\documentclass{pnastwo}

\usepackage[dvips]{graphicx}
\usepackage[tight,footnotesize]{subfigure}
\usepackage{mathrsfs}
\usepackage{longtable}
\usepackage{url}
\usepackage{physics}

\usepackage{verbatim}
\usepackage{xspace}
\usepackage{float}
\usepackage{perpage}
\usepackage{morefloats}

\usepackage[english]{babel}
\usepackage{multirow}
\usepackage{dcolumn}
\usepackage{bm}
\usepackage{listings}
\usepackage{lineno}
\usepackage{color}

\usepackage{lmodern}

\graphicspath{{figures/}}

\usepackage{amssymb,amsfonts,amsmath}

\contributor{~~}
\copyrightyear{~~}
\issuedate{~~}
\volume{~~}
\issuenumber{~~}

\begin{document}


\title{Recent Progress in Search for Dark Sector Signatures}





\author{
M.A.~Deliyergiyev
\affil{1}{Department of Experimental Particle Physics, Jozef Stefan Institute, Jamova 39, SI-1000 Ljubljana, Slovenia}
\affil{2}{Department of High Energy Nuclear Physics, Institute of Modern Physics, CAS, Nanchang Road 509, 730000 Lanzhou, China}
}
\contributor{Submitted to Open Physics}

\maketitle

\begin{article}

\begin{abstract}
Many difficulties are encountered when attempting to pinpoint a common origin for several observed astrophysical anomalies, and when assessing their tension with existing exclusion limits. 
These include systematic uncertainties affecting the operation of the detectors, our knowledge of their response, astrophysical uncertainties, and the broad range of particle couplings that can mediate interaction with a detector target. 
Particularly interesting astrophysical evidence has motivated a search for dark-photon, and focused our attention on a Hidden Valleys model with a GeV-scale dark sector that produces exciting signatures. Results from recent underground experiments are also considered. \\
There is a `light' hidden sector (dark sector), present in many models of new physics beyond the Standard Model, which contains a colorful spectrum of new particles. 
Recently, it has been shown that this spectrum can give rise to unique signatures at colliders when the mass scale in the hidden sector is well below a TeV; as in Hidden Valleys, Stueckelberg extensions, and Unparticle models. These physics models produce unique signatures of collimated leptons at high energies. By studying these ephemeral particles we hope to trace the history of the Universe. Our present theories lead us to believe that there is something new just around the corner, which should be accessible at the energies made available by modern colliders.
\end{abstract}


\abbreviations{12.60-.i; 13.85.Qk; 14.70.Bh; 14.70Pw; 14.80Bn}





\section{Introduction}

Over the last few decades, cosmology has turned into an experimental science, with more and more high-quality data becoming available.  The most surprising conclusion, based on cosmic microwave background (CMB) anisotropies -- the relic radiation of the Big Bang that fills the Universe, measured by the WMAP satellite \cite{Bennett_Astrophys_JSuppl} and based on the power spectrum of galaxy density fluctuations measured by the SDSS collaboration \cite{K.Abazajian_SDSS_Collaboration,M.Tegmark_SDSS_Collaboration} -- is that most of matter in the universe is not directly observable. 
There is a very good reason to expect that this ``extra matter'' is not normal matter. It was well established from the first phase of precision measurements that content of the Universe has to include interesting and not yet fully understood components. These are dark matter (DM), a strange matter field which interacts only gravitationally with the known matter, and dark energy (DE), an exotic component of the Universe with negative pressure which causes the accelerated expansion of the Universe. 

The observed accelerated expansion of the Universe is considered the main mystery in modern cosmology and one of the major issues confronting physics at the beginning of the new millennium. For a substance that is utterly invisible, DM does a remarkably good job of making its presence felt. The supporting evidence for the existence of DM starts to show up since the 1930's with various astrophysical measurements. Currently we can't say with certainty that dark matter has been created in the lab or directly observed. Instead, its existence has been inferred from the binding of galaxies in clusters, the rotation curves of galaxies, large scale structure simulations and observations from high-$z$ supernovae, colliding clusters of galaxies, gravitational lensing and so on \cite{RevModPhys.84.1127, Apunevych}.

Over the last 15 years we have established with very high confidence that the matter density (visible and dark) accounts for only $27\%$ of the critical value $\Omega_{M}+\Omega_{\Lambda}\approx$ 1\footnote{The total mass-energy density of the Universe. The first, $\Omega_{M}$, is a measure of the present mean mass density in nonrelativistic matter, mainly baryons and nonbaryonic dark matter. This also includes a measure of the present mass in the relativistic 3-K thermal cosmic microwave background radiation, which almost homogeneously fills space, and the accompanying low-mass neutrinos. $\Omega_{\Lambda}$ is the Dark Energy density.}, so the rest is attributed to the vacuum energy density, and thus to the cosmological constant. Moreover, we have concluded from CMB data that the Universe has a dark, or vacuum, energy density $\Omega_{\Lambda}$, of 0.73, which seems to be eerily uniform -- smoothly distributed through space, and persistent (non-diluting) through time. While the dark matter density, $\Omega_{DM}$, comprises 0.23 of the total matter-energy density of the Universe, the density of ordinary (baryonic) matter observed in the laboratory, $\Omega_{b}$, is 0.04 (only $0.5\%$ is visible using optical astronomy, with another $0.5\%$ visible at X-ray wavelengths) \cite{Bennett_Astrophys_JSuppl, Cosmic_Baryon_Budget}. These numbers tie in with the data from optical observations -- a powerful confirmation that the underlying models are correct. 
Nonetheless, for some reason, most of the matter of the Universe is stored in invisible form.

It seems that the Universe is a more complex than we suspect, and we should work to find evidence for a more complex theory of dark matter, called the $dark~sector$, comprising at least two different components: DM and DE. So, there is at least that much structure in the dark sector -- some form of energy density that can be reliably inferred through the gravitational fields it creates, but which we haven't been able to make or touch directly ourselves. This sector could include multiple types of DM and a number of dark forces, which, like ordinary matter, could combine to form some dark atoms \cite{Glashow:2005jy, Fargion:2005ep, Belotsky:2006pp, Kaplan:2009de}. 
One of the most interesting features of these proposed models is the presence of both neutral and ionized dark matter components \cite{Kaplan:2009de}. 
This idea is being tested in an experiment called the A Prime Experiment (APEX)\cite{APEX_Beacham:2013ka}.

A plethora of models of the dark sector have been proposed in order to resolve some problems in the Standard Model of particle physics. Most of them consider DE and DM as non-interacting fields, but none is really appealing on theoretical grounds. More recent models include interaction in the dark sector, and an interesting analysis of the viability of such an interaction  \cite{D.Comelli_PhysLettB_2003}. The effects of a possible dark interaction on the dynamics of galaxy clusters appear to be in agreement with observational data \cite{Abdalla_PhysLettB_2003}.

But, so far, there's no evidence of anything interesting beyond that. Indeed, the individual components of DM and DE seem relatively vanilla and featureless; more precisely, taking them to be ``minimal'' provides an extremely good fit to the data. For DM, ``minimal'' means that the particles are cold (slowly moving) and basically non-interacting with each other. For dark energy, ``minimal'' means that it is perfectly constant throughout space and time -- a pure vacuum energy, rather than something more dynamic.

Still -- all we have at hand so far are upper limits, not firm conclusions. It is certainly possible that there is a bushel of interesting physics going on in the dark sector, but it's just too subtle for us to have noticed yet. Therefore, it's important for the theorists to propose specific, testable models of a non-minimal dark sector, so that observers have targets to shoot for while they try to constrain just how interesting the darkness really is.
\begin{figure*}
  \centering
\includegraphics[scale=0.89]{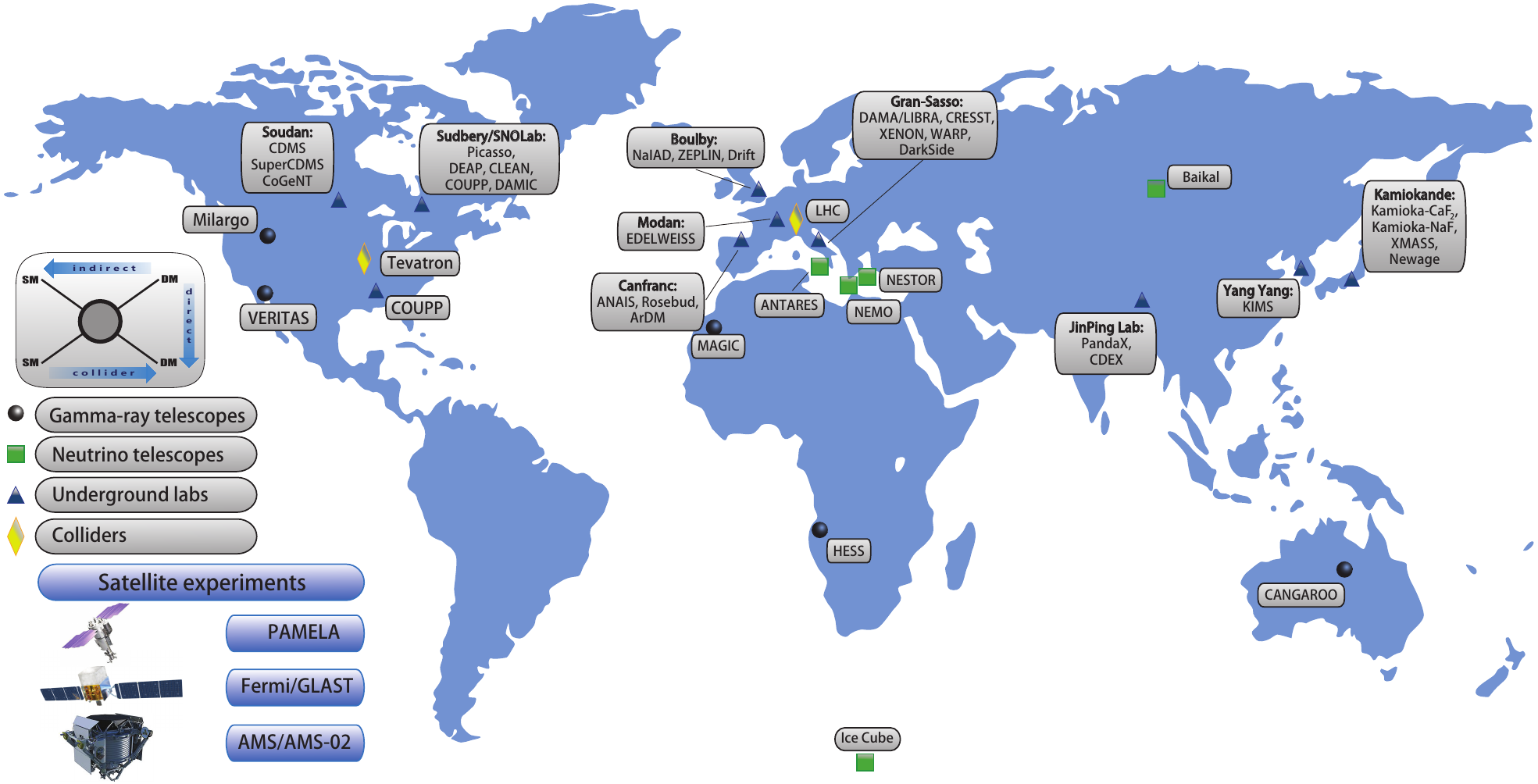}
 \caption{
Geographical location of the terrestrial experimental sites hunting for dark matter, indicating which have now stopped operation and which are still operational. The diagram shows how dark matter searches rely on interactions between DM and SM particles. Gamma-ray telescopes (black circle), neutrino telescopes (green box) -- indirect searches, which have tried to catch some bumps in the SM particle spectrum in cosmic rays with ground based or space observatories. The underground labs (magenta pyramid) -- direct searches: DM-SM scattering, looking for the nuclear recoil from galactic DM scattering, due to very weak signals the experiments are located deeply underground. Colliders (yellow diamond) -- production of the DM particles via SM interactions. Also, satellite experiments are shown (see text), which continue to explore the nature of DM.
}
  \label{fig:DarkMatter_ExprGlobalMap}
\end{figure*}

\section{Direct and indirect search for dark matter}
\label{Direct and indirect search}

Various direct and indirect experiments have been conducted in the search for dark matter (see Fig.\ref{fig:DarkMatter_ExprGlobalMap}). 
Most of the attempts to detect DM directly have started from the assumption that the DM is a haze of weakly interacting, massive particles (WIMPs) \cite{PhysRevLett.39.165}, in the mass range 10 GeV $\textless M_{\chi} \textless$ few TeV, left over from thermal or non-thermal relics created in the early universe. 
These may constitute part or all of the observed DM, $\Omega_{DM}h_{0}^{2}\sim 0.1$. This remarkable coincidence has been termed the ``WIMP miracle''\footnote{For $(m_{X}, g_{X})$ $\sim$ $(m_{weak}, g_{weak})$, the relic density is typically within an order of magnitude of the observed value}, and is perhaps the most compelling reason why our theoretical and experimental efforts are focused on searching for DM at the weak scale.

The observed DM abundance and the spectrum of its density inhomogeneities could naturally be produced by inflation if the DM particle were to be an ultra-light vector boson \cite{Graham:2015rva}. This fundamental and widely expected connection between particle physics and cosmology is the driving motivation behind most DM searches. The `massive' part would explain gravity, and the `weakly interacting' part would explain invisibility: the WIMPs would fly through stars, planets and people in untold numbers, almost never interacting. These particles also are very promising in terms of direct and indirect detection, because they must have some connection to Standard Model particles. 
Indirect detection is particularly attractive in this respect. Cosmic ray detectors, such as The Payload for Antimatter Matter Exploration and Light-nuclei Astrophysics (PAMELA) \cite{Adriani_PAMELA_PhysRevLett_2008}, The Advanced Thin Ionization Calorimeter (ATIC) \cite{J.Chang_ATIC2005}, Fermi/GLAST \cite{Fermi_GLAST} and the Alpha Magnetic Spectrometer (AMS-01, AMS-02) \cite{Aguilar2007145, PhysRevLett.110.141102}, which could detect DM indirectly by recording an anomaly in the flux of cosmic rays, have the prospect of detecting DM, if DM annihilates to some set of Standard Model states, see diagram on Fig.\ref{fig:DarkMatter_ExprGlobalMap}.

Among the WIMP candidates are the Lightest Supersymmetric Particles (LSP), occurring in supersymmetric models, like for instance neutralinos, heavy neutrino-like particles, Kaluza-Klein (KK) excitations in Extra-Dimensional theories, lightest T-odd particles in little Higgs models and other, perhaps more exotic, proposals. SuperWIMPs interact with less strength than WIMPs \cite{Feng:2003xh}, gravitationally or otherwise, and can be also good DM candidates. Among these are the gravitino, the axino (superpartner of the axion), KK gravitons and so on (for a review see \cite{Nath2010185, Baer:2008uu}).

\begin{figure*}[ht]
  \centering
\subfigure[]{
\includegraphics[scale=0.7]{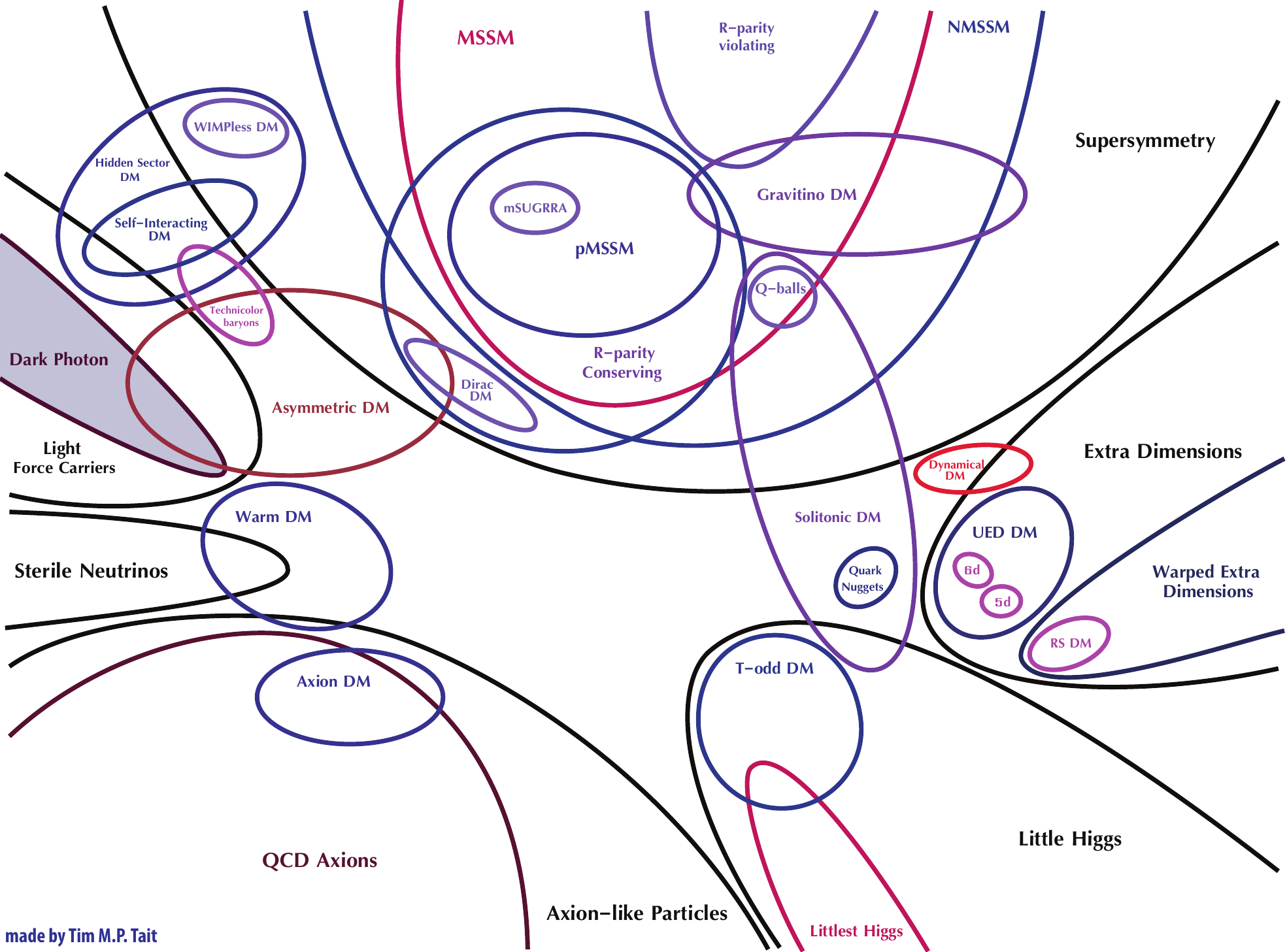}
}
 \caption{
Illustration of the sea of various dark matter candidates that considered in the literature (from Tim M. P. Tait, see \cite{Feng:2014uja}).
}
  \label{fig:SM_to_DM_particles}
\end{figure*}

Non-WIMP candidates are the WIMPzillas \cite{WIMPzillas:1998}, Cryptons \cite{Cryptons_PhysRevD.59.047301}, $Q$-balls (non-topological solitons) \cite{Coleman1985263, Kusenko:1997, Kusenko:1998}, Black Hole remnants \cite{Rajagopal1991447, PhysRevLett.85.5042} and other very massive astrophysical objects, as well as moduli fields\footnote{In quantum field theory, the term moduli fields is sometimes used to refer to scalar fields whose potential energy function has continuous families of global minima. In String Theories this term is used to refer to various continuous parameters which label possible string backgrounds, or to parameters that control the smooth deformations of shape and size of the manifold.} in String Theories or other proposed massive objects or particles Fig.\ref{fig:SM_to_DM_particles}, that are not part of the Standard Model (for a review see \cite{Nath2010185, Baer:2008uu}).

Other non-WIMP candidates include, for example, the axion \cite{axion_PhysRevLett.38.1440, axion_PhysRevD.16.1791, axion_PhysRevLett.40.223, Rusov:2013uaa}, in the mass range $10^{-5}$ - $10^{-3}$ eV \cite{axions_doi:10.1146/annurev.nucl.56.080805.140513}. The lower bound produces too high a relic density and the upper bound stems from limits on stellar cooling. 
The axion is a perfect well motivated cold dark-matter (CDM) candidate which couples weakly to two photons, moreover, this particle can explain the Sun luminosity mechanism\footnote{Mechanism that responsible for the total power emitted by the Sun.}, the extragalactic background light and may be associated with many geophysical observations \cite{Rusov:2013uaa}. Ongoing experiments like ADMX \cite{axions_PhysRevD.69.011101}, and CAST \cite{axions_1475-7516-2007-04-010} have put limits on its coupling to photons and its mass but its discovery is still lacking. 


Other non-WIMP candidates include, for example, the axion \cite{axion_PhysRevLett.38.1440, axion_PhysRevD.16.1791, axion_PhysRevLett.40.223, Rusov:2013uaa}, in the mass range $10^{-5}$ - $10^{-3}$ eV \cite{axions_doi:10.1146/annurev.nucl.56.080805.140513}. The lower bound produces too high a relic density and the upper bound stems from limits on stellar cooling. The axion is a reasonable and accepted cold dark-matter (CDM) candidate which couples weakly to two photons, moreover, this particle can explain the Sun luminosity mechanism\footnote{Responsible for the total power emitted by the Sun.}, the extragalactic background light, and may be associated with many geophysical observations \cite{Rusov:2013uaa}. Ongoing experiments like ADMX \cite{axions_PhysRevD.69.011101}, and CAST \cite{axions_1475-7516-2007-04-010} have put limits on its coupling to photons and on its mass, but its discovery is still lacking.

Several experiments are currently searching for WIMPs via their elastic scattering interactions with nuclei, one of which is the Cryogenic Dark Matter Search (CDMS) \cite{Akerib:2005zy} in the disused Soudan Iron Mine in northern Minnesota. One usually assumes that the WIMP scatters elastically, and that the spin-independent cross section for scattering off protons and neutrons is roughly the same. At cryogenic temperatures of just 40 milikelvin the simultaneous measurement of phonon and ionization signals in semiconductor detectors permits event by event discrimination between nuclear and electronic recoils down to 5 to 10 keV recoil energy, so that any heat associated with a WIMP impact could be detected. 
That team is now running a second-generation experiment, called CDMSII \cite{Ahmed:2008eu}, which aims to increase the sensitivity by a factor of 10.
\begin{figure*}[ht]
\centering
\subfigure[]{
\includegraphics[scale=0.26]{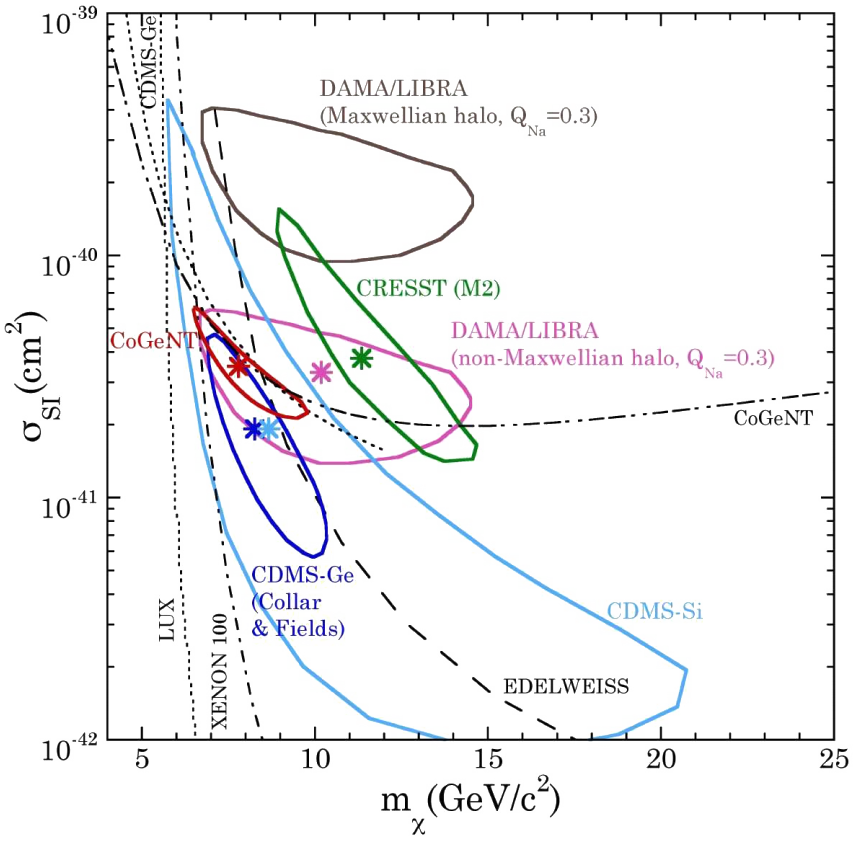}
\label{fig:GoGENT_results2014}
}
\subfigure[]{
\includegraphics[scale=0.85]{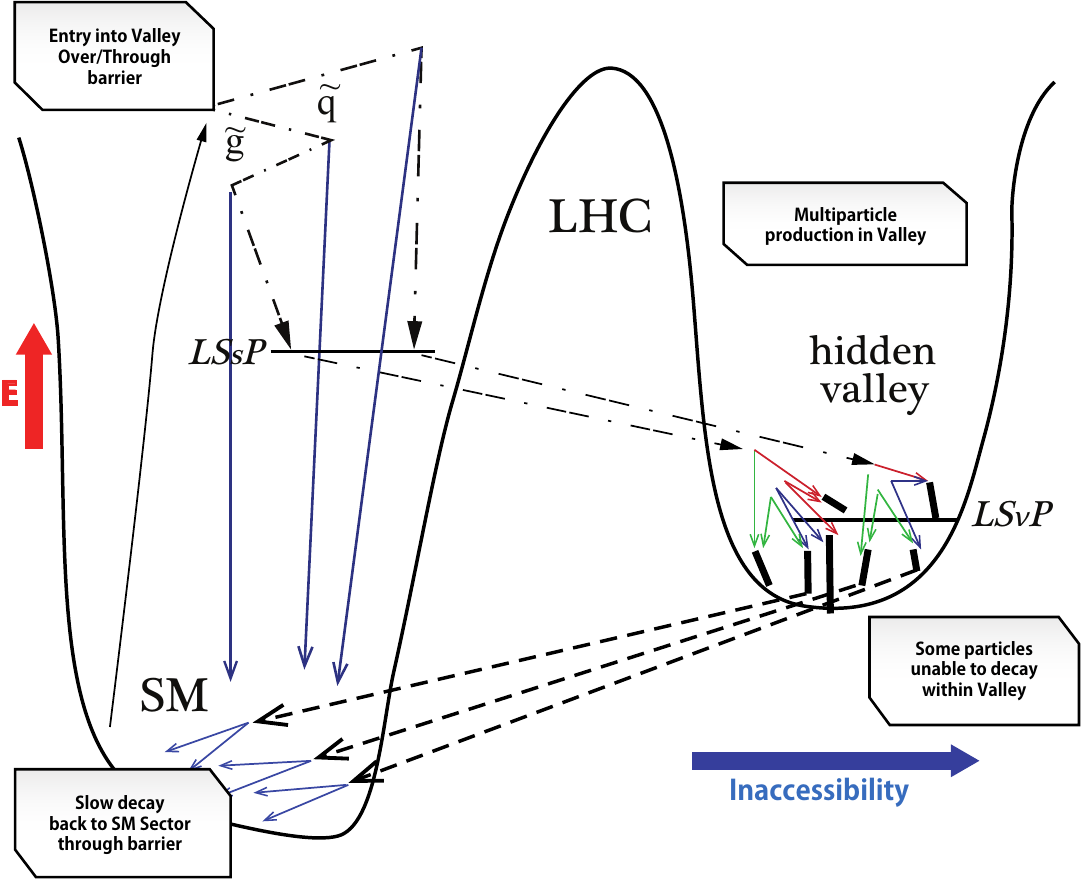}
\label{fig:SchematicOf_TheHiddenValley_type}
}
\caption{
Left $\textendash$ the effect on the DAMA/LIBRA ROI when a factor 6.8 larger fractional modulation than that predicted by the SHM \cite{Freese:2012CO} is assumed, i.e., the same as that found in the CoGeNT dataset (free $T_{1/2}$ case): a nonMaxwellian local halo favoring large values of $S$ is capable of reconciling the tension between DAMA/LIBRA and other recently reported anomalies, providing a coherent picture for these observations. (Figure from \cite{Aalseth:2014eft}). 
Right $\textendash$ a schematic of the Hidden Valley type dark sectors under consideration, where  each superpartner decays to hard jets/leptons and  the lightest standard model  superpartner (LSsP); the LSsP then decays to an LSvP plus other v-hadrons, some of which decay to softer jet/lepton pairs. (Figure from \cite{M.J.Strassler_2006}).
}
\label{fig:Communication_VS_and_HS}
\end{figure*}

The XENON experiment \cite{PhysRevLett.100.021303}, which measures simultaneously the scintillation and the ionization produced by radiation in pure liquid xenon\footnote{Searching a similar mass range in comparison with CDMS}, and CDMS \cite{PhysRevLett.93.211301} both claim to have detected the very low-mass WIMPs, however their results contradict those from other experiments. For spin dependent WIMPs with pure neutron couplings XENON has the best published limit at all masses \cite{PhysRevLett.101.091301}. Results reported by CDMS came from its lighter silicon detectors and not its heavier germanium detectors. Since interaction between a WIMP and a nucleus should relate to the amount of nucleons in the nucleus, WIMPs should more likely to interact with germanium detectors, since germanium is much heavier than silicon. Indeed, when the CDMS collaboration designed their experiment in the early 2000's, they included the silicon ones only to verify results from the germanium. 

Very recently, the CRESST-II collaboration has announced results for their DM search with 730 kg-days of net exposure in a $\rm{CaWO}_{4}$ target \cite{Angloher:2011uu}. 
They were able to explore masses down to 0.5 GeV, a novelty in the field of direct dark matter searches \cite{Angloher:2015ewa}. This observation could not be explained by known background and is compatible with a DM signal also rising at low energies.

Nevertheless, perhaps the most intriguing, and most controversial, of these experiments is the Dark Matter Large Sodium Iodide Bulk for Rare Processes (DAMA/LIBRA)~\cite{BernabieRiccardo_2005}, which shares space with XENON and CRESST at Gran Sasso National Laboratory in Italy. DAMA works on the principle that the Sun's orbit around the center of the Galaxy carries the Solar System through the invisible cosmic background of DM at some 220 kilometers per second. The Earth orbit around the Sun implies that the amount of DM incident on the Earth varies by about $10\%$ from summer to winter. The maximum would correspond to June and the minimum to December \cite{BernabieRiccardo_2008}.

Thus, detectors on Earth should have DM flowing through them at this velocity, modulated by an annual variation of about 30 kilometers per second as the Earth orbits the Sun.

The DAMA team tries to observe the scintillation of nuclear recoil events\footnote{Detection of the scintillation light created by the elastic low energy nuclear recoils in the scattering of DM candidates with the detector material. The measurement of the decreasing ionization rate along the recoil trajectory provides the tag of the incoming DM candidates.} inside a thallium-doped sodium iodide crystal (salt) detector, and claims to have followed just such a periodic DM signal for thirteen years with a very high statistical significance of 8.9$\sigma$ in the 2-6 keV energy interval. Considering the former DAMA/NaI and the recent DAMA/LIBRA data together (total exposure 1.17 ton$\times$yr)
\cite{BernabieRiccardo_2010}
\footnote{The analogue of the total luminosity in accelerator physics. The expected events (activations) per unit of time defined as $dN/dt$ $\sim$ $\sigma_{x}\rho_{\rm{target}}\Phi/A = \frac{1}{\rm{day}}\frac{\rm{kg}}{\rm{m^{3}}}$, where $\Phi$ is the the hypothetical dark-particle's flux per square meter per day, $N$ number of expected events, $\sigma_{x}$ is an inelastic/elastic cross section of the hypothetical dark particle, and $A$ is the target mass number. In the direct detection experiments, experimentalists have resorted to an increase in the volume of the effective target, $V_{\rm{target}}$, as well as an increase in the exposure time, $t_{\rm{exposure}}$ in order to increase the potential number of observed events.}, 
this result is consistent with Earth's orbit through the Galactic DM halo. At this point we should note that the distribution of galactic DM may not be smooth; galaxy formation is an ongoing process, and computational studies suggest that there may be a significant amount of DM substructure in the form of clumps and tidal streams \cite{Helmi01032003, Ibata01062002}. Different models of the halo can shift the phase of this modulation completely, turning expected maxima into minima and vice-versa, changing the expected amplitude as well. The detector crystals cannot distinguish between WIMPs and background events from ordinary radiation in the detector's surroundings, so this result depends on the assumption that background events occur at a constant rate that does not vary with the season.

Possible background reactions have also been carefully searched for. The only process which has been found as a hypothetical possibility is the muon flux modulation reported by the MACRO experiment \cite{Ambrosio1997109}. In fact, the observed muon flux shows a nearly sinusoidal time behavior with one year period and maximum in the summer with amplitude of $2\%$; this muon flux modulation is correlated with the temperature of the Earth's atmosphere. The contribution of this effect with solar neutrinos flux is many orders of magnitude lower than the modulation amplitude measured by DAMA/LIBRA. ``Fake'' signals can be caused not only by cosmic rays, but also radioactive isotopes occurring in nature. The latter issue requires us to consider the correlations between solar activity and nuclear decay rates of the materials used in DAMA as sensors. Evidence for this unexplained periodic fluctuation has been presented in a recent series of papers; this includes an apparent correlation between Earth-Sun distance, using data taken at Brookhaven National Laboratory(BNL), and at the Physikalisch-Technische Bundesanstalt(PTB) \cite{Jenkins200942, Jenkins2010332}.

The DAMA/LIBRA result has so far not been confirmed, and will remain as such until possible corroboration by the latest data coming from the Coherent Germanium Neutrino Technology (CoGeNT) collaboration. Its detector sits near the CDMS detector in the Soudan Mine, and the team claims a detection of an annual modulation with a statistical significance of 2.7-2.8$\sigma$, which is at the margin of statistical significance of a signal in  particle physics \cite{Hooper:2011hd, Sinervo:2002sa}. The modulation is most pronounced in the 0.5-2 keV region, while it is absent for surface events.

CoGeNT is tuned to detect incoming particles with much lower masses than those sought by its neighbor, CDMS. It was originally intended to explore this range to rule out the existence of low-mass WIMPs, but its results only ended up making things murkier. Around the time that the CDMSII reported its `nearly nothing' findings, CoGeNT released data from its first 56 days of operation \cite{CoGeNTColl_PhysRevLett.106.131301}. The results showed hundreds of particle events that could be interpreted as DM with a mass between 7 and 11 GeV. 

The CoGeNT collaboration has recently published results covering more than three years of continuous operation \cite{Aalseth:2014eft}. Their conclusions have not changed: the signal is still present with about the same strength and in the same area, seasonal modulation is observed as before. However, backgrounds and/or systematic sources which could mimic an annual modulation signature in DM detectors are presently unknown.

Today the situation looks as follows: the results of the DAMA/LIBRA and CoGeNT collaborations have appeared to be in conflict with the limits from other experiments including XENON and CDMS. The parameter region consistent with the CoGeNT signal can be excluded by CDMS, XENON10 and XENON100, see Fig.\ref{fig:Communication_VS_and_HS}(a). One would have to assume that these three experiments are incorrect in terms of energy scale in order to reconcile their limits with the CoGENT signal. On the other hand, the observation made by CoGENT is very intriguing, especially in combination with the long-standing DAMA modulation signal and the oxygen band excess in CRESST \cite{Angloher:2011uu, Angloher:2015ewa}. Furthermore, due to the fact that different collaborations use different substances, having different thresholds, the disagreement between them is not so unreconcilable. The Ref.\cite{Aalseth:2014eft} provides an analysis of the CoGeNT data under different models of the distribution of DM in the galaxy, and demonstrates that there is an option in which three experiments with a positive result (DAMA, CoGeNT, CDMS) are consistent with each other, see Fig.\ref{fig:Communication_VS_and_HS}(a). There, of course, a problem remains, as understood in the light of the recent negative results of the experiment LUX \cite{Akerib:2013tjd}, but at least this point of criticism is itself not indisputable.

The aforementioned detection strategies are appropriate if DM candidates are relatively massive, with low phase-space density and particle-like behavior. However, another generic class of DM candidates exhibits field-like behavior. A light (bosonic) field making up the DM with mass $\ll$ 0.1 eV will have a high phase-space density, since the local DM density is $\rho_{DM} \sim 0.3~{\rm{GeV/cm^{3}}} \sim {(0.04~{\rm{eV}})}^{4}$. Such light-field DM is generally produced nonthermally (unlike the WIMP), often by the misalignment mechanism \cite{Graham:2015rva}, and is best described as a classical, background field oscillating with frequency roughly equal to its mass \cite{PRESKILL1983127, Abbott1983133, DINE1983137}. To search for this field-like behavior of DM candidates a new class of experiments has been proposed \cite{Jaeckel:2007ch, Horns:2012jf, Chaudhuri:2014dla}, a bit similar to those already used in the detection of the ultra high energy cosmic rays with the help of radio techniques \cite{Filonenko:2015}.

Among the most notable detection strategies in nonaccelerator type experiments are the laser polarization experiments such as BFRT \cite{PhysRevD.47.3707}, PVLAS \cite{Zavattini:2005tm}, and Q\&A \cite{Chen:2006cd}\footnote{Laser polarization experiments where linearly polarized laser light is sent through a transverse magnetic field, and changes in the polarization state are searched for.}, the ``light-shining-through-wall'' experiments (LSW) \cite{Ahlers:2007qf}, such as BFRT \cite{Ruoso:1992}, ALPS \cite{Ehret:2009sq}, and the conversion experiments from the solar DM particle flux, ``helioscopes'' \cite{Redondo:2008aa, An:2013yua, An:2014twa}.

As can be seen from the discussion above, direct detection of DM is now a mini-industry, with many experiments using a wide variety of techniques, both currently and planned. It seems that the race to detect DM has yielded mostly confusion, and we are still struggling to answer the `What is it?' question, often feeling like we're chasing a ghost. Certainly, our detectors have been giving us a lot of strange and contradictory results. Fortunately, this confusion is likely to be temporary. DM detectors are roughly 1000 times more sensitive to ultra-rare events than they were 20 years ago and that should increase by another factor of 100 over the next decade, as physicists build bigger detectors and become more skilled at suppressing the background noise that can be confused with genuine signals \cite{Andrzej:2012, Behnke:2013sma, Akerib:2015cja,  Me3u:2015}. 
An issue is that the experiments with very large target masses, with sensitivities to DM scattering cross sections of $10^{-46}-10^{-48}$  $\rm{cm}^{2}$, depending on the DM candidate mass, could not eliminate the coherent neutrino scattering, which represents an irreducible background for these searches \cite{Monroe:2007xp, Cushman:2013zza, Grothaus:2014hja}. Nevertheless, those larger, more sensitive detectors, which have already begun to operate in recent years, could soon change that picture.  The Large Hadron Collider (LHC), the world's highest-energy particle accelerator at CERN, Europe's high-energy physics lab near Geneva, Switzerland, which is joining the DM pursuit, is now pushing high energy physics into new territory and presents an unprecedented opportunity to probe the realm of new physics in the TeV region and shed light on some of these core unresolved issues of Particle Physics. These include the possible constituents of DM, i.e. the possible existence of extra gauge groups.

\section{Cosmic positron puzzle}
\label{INTRO_CosmicPositronPuzzle}

In addition to account the recent proliferation of astrophysical anomalies, like a the annual modulation of the DM signal observed in DAMA/NaI and DAMA/LIBRA \cite{BernabieRiccardo_2010}, also spotted by CoGeNT collaboration \cite{Hooper:2011hd},
there is also an observed unexpected increase in the ratio of positrons to electrons in 10-100 $\GeV$ cosmic rays within 1 kpc measured by PAMELA \cite{Adriani_PAMELA_PhysRevLett_2008, M.Boezi_IDM2008, Adriani_PAMELA_Nature_2008},
ATIC \cite{J.Chang_ATIC2005, J.Chang_Nature2008},
HESS \cite{Aharonian:2009ah},
INTEGRAL \cite{Strong_Astrophys_2005, Finkbeiner_PhysRev_2007}, and  
AMS-02 \cite{PhysRevLett.110.141102} experiments, which demonstrate the ratio of the number of positrons to the total number of electrons plus positrons, at energies between 0.5 and 350 $\GeV$. The range of the reported positron fraction extends beyond the reach of previous experiments flown on high-altitude balloons (HEAT \cite{Heat_1538-4357-482-2-L191}), or space shuttles and satellites (AMS-01 \cite{Aguilar2007145}).
Furthermore, we have to consider results coming from the Large Area Telescope (LAT) on the Fermi Gamma-ray Space Telescope, which demonstrates that the positron fraction continues its surprising increase with energy \cite{PhysRevLett.108.011103}. Fermi LAT Collaboration measured the combined electron and positron spectrum from 7 $\GeV$ to 1 $\TeV$, while PAMELA measured just the electron spectrum from 1 to 625 $\GeV$. 
Both Fermi LAT and AMS experiments agreed with the first PAMELA results, though neither measured positrons separately. Recently, the Fermi LAT Collaboration has also measured both the electron and positron spectra from 20 to 200 $\GeV$ separately, including the first measurement of the absolute positron spectrum above 50 $\GeV$ and the first determination of the positron-to-electron ratio above 100 $\GeV$. 
The Fermi LAT doesn't have a permanent magnet located at the core of the instrument in order to deflect particles tracks (charge separation) in comparison with AMS-02, however Fermi LAT Collaboration, with non-trivial adjustments in analysis, has also been able to measure the positron fraction. These adjustments is the reason why  Fermi LAT has such large uncertainties in its positron spectrum.

In addition to Fermi LAT, the new AMS results agree beautifully with what PAMELA observed, thus reinforcing the indications that the positron fraction rises with energy, but this time, with unprecedented statistics and background controls. There is however an apparent discrepancy between PAMELA and AMS-02 data below about 2 $\GeV$, perhaps due to different effects of the ``solar modulation'' on particle populations for the PAMELA and AMS-02.

The meaning of these observations is as follows: Outer space is filled with a huge amount of ordinary particles (electrons, protons, nuclei), and they are easily accelerated to high energies in a variety of ``cosmic accelerators''. In contrast, anti-particles, in particular positrons, are always suppressed. They have a very short lifetime in space, because sooner or later they stumble upon particles of matter and annihilate. Therefore, the presence of anti-particles means that their continuous production has been ``established'' in space, and there are various scenarios at this point.

Positrons can, for instance, be products of collisions of accelerated protons or electrons that happened somewhere in the space. A characteristic feature of all these processes is the smooth and steady decrease in the number of particles with energy. But it may also be that elusive DM particles can annihilate into, say, an electron-positron pair. Note, that at this point most of the models assume that the DM particles may interact with each other \cite{D.Comelli_PhysLettB_2003, Tulin:2012wi, Tulin:2013teo, Kaplinghat:2013xca, Kaplinghat:2015gha}. On the other hand DM particles can decay into electron-positron pairs. Mass of DM particles should be a substantial (and the prediction of many theories lies in the area of hundreds of $\GeV$ to several $\TeV$), and the speed of these particles should be small. Therefore, the electrons and positrons in such a collision will be produced with a more or less definite energy.  In other words, one would observe a peak in the energy distribution. This peak is quite difficult to notice in the absolute value of the flow, but if we build the ratio of positrons to the sum of electrons and positrons (at a given energy), all features will became more explicit.

Are these results evidence in favor of dark matter? Not at all, no more than the PAMELA results. We could make here even more clear statement: it would be wrong to say that obtained results give us at least some new information about DM. 
Yes, recent experiments observe the positron excess, no doubt there, but there is still no clear indication of its origin. Even if the data lie on a beautiful bump in the center area of $0.5-1 ~\TeV$, it still cannot be explained by the annihilation of DM. 
Mainly because it is required that the annihilation cross section should be several orders of magnitude higher than any acceptable values (from the point of view of other manifested and unmanifested DM particles). 
But theoreticians have already come up with some options, which may resolve inconsistencies of this cross-section with other cosmological data. For example, the factor for enhancement of the annihilation cross section via a mechanism first described by Sommerfeld \cite{Sommerfeld_ANDP:ANDP19314030302}. Also, there are models of decaying DM \cite{Bajc:2010qj}, which seem to have been able to describe the PAMELA and ATIC data.

Moreover, we have to mention that there are no clear signs of DM in gamma rays \cite{Bringmann2012194}. If DM annihilation or decay is the source, one needs to find a mechanism explaining why it populates cosmic electrons only. On the other hand, an incomplete understanding of how the cosmic rays propagate, or other astrophysical sources, could have explained the positron excess \cite{Roberto:2010, Burch:2010}. It turns out that secondary electrons have a very flat equilibrium spectrum, which is responsible for the observed positron excess \cite{SNRemnants_Blasi:2009hv}. The origin of the rising positron fraction at high energy is unknown and has been described to a variety of mechanisms including pulsars \cite{Stefano:2012, 0004-637X-772-1-18, Venter:2015oza}, cosmic rays interacting with giant molecular clouds, and DM -- ``primary'' particles, the ``secondaries'' come from these particles colliding with interstellar gas and producing pions and muons, which decay into electrons and positrons. A third, interesting possibility is that electrons and positrons are created by the annihilation of DM particles in the Milky Way and its halo, see \cite{doi:10.1146/annurev-astro-081710-102528, doi:10.1142/S0218271810018268} for recent reviews. The modulation in DAMA \cite{BernabieRiccardo_2004, Bernabei:2013xsa} is present only in the single-hit events, while it is absent in the multiple-hits as expected for the DM signal\footnote{The DM annual modulation signature is very distinctive since the effect induced by DM particles must simultaneously satisfy all the following requirements: the rate must contain a component modulated according to a cosine function (1) with one year period (2) and a phase that peaks roughly $\simeq$ 2 June (3); this modulation must only be found in a well-defined low energy range, where DM particle induced events can be present (4); it must apply only to those events in which just one detector of many actually ``fires'' (single-hit events), since the DM particle multi-interaction probability is negligible (5); and etc. Please read the Ref.\cite{Bernabei:2013xsa} for more details.}. However, a new source of leptons in the sky may also imply new sources of leptons in colliders. So, we look forward to new data from all experimental groups which are engaged in the dark matter search.

Theorists had hoped that the LHC would open the door to the dark-sector. The LHC aims to produce particles in the center-of-mass energy up to 14 TeV -- above those that are already known from previous collider experiments. At the same time, there are exciting possibilities for new physics in the low-mass range that may have gone unnoticed until now. Many models of New Physics beyond the Standard Model contain a $light~hidden~sector$ (dark sector) with a colorful spectrum of new particles \cite{J.T.Ruderman_T.Volansky_JHEP_02_2010}, which can be probed at the LHC.

As ordinary matter couples to a long-range force known as electromagnetism mediated by photons, the DM couples to a new long-range force, may be referred to as ``dark electromagnetism"  -- a new fundamental force in addition to the four that we already know about, mediated by particles known as $dark~photons$ or $hidden~photons$ \cite{Okun:1982xi, Goodsell:2009xc}. 
It would be the first sign of a hidden sector \cite{BernabieRiccardo_2005,HoomanDavoudiasl_2005, P.Brian_Wilczek_2006}, which could include an entire zoo of new particles, including DM, see Fig.\ref{fig:SM_to_DM_particles}. The idea of new forces acting on DM is by no means new. What's exciting about dark photons is that they are much more natural from a particle-physics perspective. Typical models of quintessence and long-range fifth forces invoke scalar fields \cite{Peebles:1987ek, PhysRevLett.80.1582}, which are easy and fun to work with, but which by all rights should have huge masses, and therefore not be very long-range at all. 
The production of dark photons in this case is dominated by the dark Higgsstrahlung, or equivalently, by the pair-production of $U(1)_{X}$  -- charged Higgs scalar fields, like $h_{d1}$ or $S_{d1}$.

The dark photon comes from a gauge symmetry, just like the ordinary photon, and its masslessness is therefore completely natural before symmetry breaking. The instability of this photon would help in detecting it indirectly -- after the dark photons have decayed into electrons and positrons or into pairs of muons \cite{N.Arkani-Hamed_N.Weiner_PhysRev_2008, N.Arkani-Hamed_N.Weiner_JHEP_2008}, see diagram on Fig.\ref{fig:DarkMatter_ExprGlobalMap}. 

However, there is a lot of the recent progresses developing theories around dark photons, especially in the parameter space where the mass of the dark photon is smaller than $\mathcal{O}$(MeV-GeV) \cite{Tulin:2012wi, Tulin:2013teo, Kaplinghat:2013xca, Kaplinghat:2015gha, Kaplinghat:2015aga}. Efforts to search for lighter hidden photons with smaller-scale lab experiments \cite{Graham:2014sha, Jaeckel:2007ch, Ehret:2010mh, Ringwald:2012hr, Horns:2012jf, Betz:2013dza, Chaudhuri:2014dla} have received somewhat less attention. But these experiments have the potential to discover hidden photons over a vast range of parameter space, extending from $\mathcal{O}({\rm{eV}})$ down to scales as low as $\mathcal{O}(10^{-18}{\rm{eV}})$. The cosmological bound on photon-dark photon oscillations for dark photon masses between $10^{-14}$ eV and $10^{-7}$ eV  was evaluated in \cite{Mirizzi:2009iz}. In this low mass range, the decay channel of dark photon to electron-positron pair is forbidden, and the decay life time of the dark photon is long enough for the dark photon itself to be a reasonable dark matter candidate. Also, dark photon with mass smaller than 1 MeV, which can be produced inside stars and the star cooling bound, can be very useful to constrain the dark photon parameter space \cite{An:2013yfc, Redondo:2013lna}. This explains efforts with DM direct detection experiments such as XENON, CoGeNT and LUX to detect the dark photons emitted from the Sun directly \cite{Redondo:2013lna, An:2013yua, An:2014twa}.

We can therefore imagine a completely new kind of photon, which couples to DM but not to ordinary matter. So there could be dark electric fields \cite{SteveReucroft:2014}, dark magnetic fields \cite{ZurabBerezhiani:2013}, dark radiation \cite{Feng:2009mn, Baek:2013dwa, Buschmann:2015awa}, etc. The dark matter itself consists half of particles with dark charge +1, and half with antiparticles with dark charge -1. At this point one may ask, ``Why don't the dark particles and antiparticles all just annihilate into dark photons?'' That kind of thinking is probably why ideas like this weren't explored twenty years ago (as far as we know). But there is clearly a range of possibilities for which the DM doesn't annihilate very efficiently; for example, if the mass of the individual DM particles were sufficiently large, their density would be very low, and they just wouldn't ever bump into each other. 
Alternatively, if the strength of the new force was extremely weak, it just wouldn't be that effective in bringing particles and antiparticles together. The strength of the dark electromagnetic force is characterized, naturally, by the dark fine-structure constant; remember that ordinary electromagnetism is characterized by the ordinary fine-structure constant $\alpha$ =1/137. However, we know a little more about the DM than ``it doesn't annihilate''. We also know that it is close to collisionless -- DM particles don't bump into each other very often. If they did, all sorts of things would happen to the shape of galaxies and clusters that we don't actually observe \cite{Tulin:2012wi, Tulin:2013teo, Kaplinghat:2013xca, Kaplinghat:2015gha, Kaplinghat:2015aga}. So there is another limit on the strength of dark electromagnetism: interactions should be sufficiently weak that DM particles don't ``cool off'' by interacting with each other in galaxies and clusters. That turns into a more stringent bound on the dark fine-structure constant: about an order of magnitude smaller, at $\hat{\alpha}$ $\textless$ $10^{-3}$. It has been shown that models with electrically charged DM in which $\alpha = \hat{\alpha}$ are ruled out unless the mass of the DM particle is heavier than a few TeV. In order to evade these constraints and build an unbroken dark $U(1)$ with correct relic abundance, one may need to introduce a short-range force coupling which should be relatively small in order to evade Galactic dynamics bounds \cite{Ackerman:2008gi}.

\section{Cosmic positron puzzle}
\label{INTRO_CosmicPositronPuzzle}

In addition to account the recent proliferation of astrophysical anomalies, like a the annual modulation of the DM signal observed in DAMA/NaI and DAMA/LIBRA \cite{BernabieRiccardo_2010}, and also spotted by the CoGeNT collaboration \cite{Hooper:2011hd},
there is an observed unexpected increase in the ratio of positrons to electrons in 10-100 GeV cosmic rays within 1 kpc measured by PAMELA \cite{Adriani_PAMELA_PhysRevLett_2008, M.Boezi_IDM2008, Adriani_PAMELA_Nature_2008},
ATIC \cite{J.Chang_ATIC2005, J.Chang_Nature2008},
HESS \cite{Aharonian:2009ah},
INTEGRAL \cite{Strong_Astrophys_2005, Finkbeiner_PhysRev_2007}, and  
AMS-02 \cite{PhysRevLett.110.141102} experiments, which demonstrate the ratio of the number of positrons to the total number of electrons plus positrons, at energies between 0.5 and 350 GeV. The range of the reported positron fraction extends beyond the reach of previous experiments flown on high-altitude balloons (HEAT \cite{Heat_1538-4357-482-2-L191}), or space shuttles and satellites (AMS-01 \cite{Aguilar2007145}).
Furthermore, we have to consider results coming from the Large Area Telescope (LAT) on the Fermi Gamma-ray Space Telescope, which demonstrates that the positron fraction continues its surprising increase with energy \cite{PhysRevLett.108.011103}. The Fermi LAT Collaboration measured the combined electron and positron spectrum from 7 GeV to 1 TeV, while PAMELA measured just the electron spectrum from 1 to 625 GeV. Both Fermi LAT and AMS experiments agreed with the first PAMELA results, though neither measured positrons separately. Recently, the Fermi LAT Collaboration has also measured both the electron and positron spectra from 20 to 200 GeV separately, including the first measurement of the absolute positron spectrum above 50 GeV and the first determination of the positron-to-electron ratio above 100 GeV. The Fermi LAT doesn't have a permanent magnet located at the core of the instrument in order to deflect particle tracks (charge separation) in comparison with AMS-02, however Fermi LAT, with non-trivial adjustments in analysis, has still been able to measure the positron fraction. These adjustments do however lead to the large uncertainties in the Fermi LAT positron spectrum.

In addition to Fermi LAT, the new AMS results agree beautifully with what PAMELA observed, thus reinforcing the previous indications that the positron fraction rises with energy, but this time with unprecedented statistics and background controls. There is however an apparent discrepancy between PAMELA and AMS-02 data below about 2 GeV, perhaps due to different effects of the ``solar modulation'' on particle populations for the PAMELA and AMS-02.

The meaning of these observations is as follows: Outer space contains a huge amount of ordinary particles (electrons, protons, nuclei), and they are easily accelerated to high energies in a variety of ``cosmic accelerators''. In contrast, anti-particles, in particular positrons, are always suppressed. They have a very short lifetime in space, because sooner or later they stumble upon particles of matter and annihilate. Therefore, the presence of anti-particles means that their continuous production has been ``established'' in space, and there are various scenarios at this point.

Positrons can, for instance, be products of collisions of accelerated protons or electrons that happened somewhere in space. A characteristic feature of all these processes is the smooth and steady decrease in the number of particles with energy. But it may also be that elusive DM particles can annihilate into, say, an electron-positron pair. Note that at this point most of the models assume that the DM particles may interact with each other \cite{D.Comelli_PhysLettB_2003, Tulin:2012wi, Tulin:2013teo, Kaplinghat:2013xca, Kaplinghat:2015gha}. On the other hand DM particles can decay into electron-positron pairs. The mass of DM particles should be a substantial (Indeed many theoretical predictions lie in the area of hundreds of GeV to several TeV), and the speed of these particles should be small. Therefore, the electrons and positrons in such a collision will be produced with a more or less definite energy.  In other words, one would observe a peak in the energy distribution. This peak is quite difficult to notice in the absolute value of the flow, but if we build the ratio of positrons to the sum of electrons and positrons (at a given energy), all features will became more explicit.

Are these results evidence in favor of dark matter? Not at all, no more than the PAMELA results. We could make here even more clear statement: it would be wrong to say that obtained results give us at least some new information about DM. There is no doubt that recent experiments observe the positron excess, but there is still no clear indication of its origin. Even if the data lie on a beautiful bump in the center area of $0.5-1$ TeV, this still cannot be explained by the annihilation of DM. Mainly because it is required that the annihilation cross section should be several orders of magnitude higher than any acceptable values (from the point of view of other manifested and unmanifested DM particles). 
But theoreticians have already come up with some options, which may resolve inconsistencies of this cross-section with other cosmological data. For example, the factor for enhancement of the annihilation cross section via a mechanism first described by Sommerfeld \cite{Sommerfeld_ANDP:ANDP19314030302}. Also, there are models of decaying DM \cite{Bajc:2010qj}, which seem to have been able to describe the PAMELA and ATIC data.

Moreover, we have to mention that there are no clear signs of DM in gamma rays \cite{Bringmann2012194}. If DM annihilation or decay is the source, one needs to find a mechanism explaining why it populates cosmic electrons only. On the other hand, an incomplete understanding of how the cosmic rays propagate, or other astrophysical sources, could have explained the positron excess \cite{Roberto:2010, Burch:2010}. It turns out that secondary electrons have a very flat equilibrium spectrum, which is responsible for the observed positron excess \cite{SNRemnants_Blasi:2009hv}. The origin of the rising positron fraction at high energy is unknown and has been explained by a variety of mechanisms including pulsars \cite{Stefano:2012, 0004-637X-772-1-18, Venter:2015oza}, cosmic rays interacting with giant molecular clouds, and DM  ``primary'' particles; the ``secondaries'' come from these particles colliding with interstellar gas and producing pions and muons, which decay into electrons and positrons. A third, interesting possibility is that electrons and positrons are created by the annihilation of DM particles in the Milky Way and its halo, see \cite{doi:10.1146/annurev-astro-081710-102528, doi:10.1142/S0218271810018268} for recent reviews. The modulation in DAMA \cite{BernabieRiccardo_2004, Bernabei:2013xsa} is present only in the single-hit events, while it is absent in the multiple-hits as expected for the DM signal\footnote{The DM annual modulation signature is very distinctive since the effect induced by DM particles must simultaneously satisfy all the following requirements: the rate must contain a component modulated according to a cosine function (1) with one year period (2) and a phase that peaks roughly $\simeq$ 2 June (3); this modulation must only be found in a well-defined low energy range, where DM particle induced events can be present (4); it must apply only to those events in which just one detector of many actually ``fires'' (single-hit events), since the DM particle multi-interaction probability is negligible (5); and etc. Please read the Ref.\cite{Bernabei:2013xsa} for more details.}. However, a new source of leptons in the sky may also imply new sources of leptons in colliders. So, there are many reasons to look forward to new data from all experimental groups which are engaged in the dark matter search.

\section{Constraints on dark sector from colliders}
\label{Constraints on dark sector from colliders}

The aforementioned astrophysical results allow us to use a model of DM which does not predict  antiproton annihilation. Actually, this is easiest done by forcing the DM particles to annihilate into leptons, avoiding decays into hadrons, but in this case these hypothetical particles will be produced in hadron collisions only very rarely. Therefore, with the hypothesis, assuming that DM annihilation may be the source of ATIC, PAMELA, Fermi LAT,  AMS-02 excesses, we can come straight to a couple of interesting conclusions. Firstly, the increase in positrons was a potential sign that DM particles had been detected: these particles annihilate to electrons or muons, either directly or indirectly; and secondly, they do so about 100 times more frequently than expected. 

Both can be accomplished with a theory containing a new force in the specific subclass of Hidden Valleys (``dark sectors"), 
with a dark vector boson mixing with the photon, if its mass is $\mathcal{O}$(GeV). In this case special attention should be paid to a model of DM by N. Arkani-Hamed et al.~\cite{N.Arkani-Hamed_N.Weiner_PhysRev_2008, N.Arkani-Hamed_N.Weiner_JHEP_2008, BaumgartMatthew_JHEP_2009}. They proposed that DM is charged under non-Abelian ``hidden''  gauge symmetry $G_D$, which is broken at $\sim$~1 GeV scale \cite{N.Arkani-Hamed_N.Weiner_JHEP_2008}. Such a model reconciles the tensions between the astrophysical experimental results and more conventional models of WIMP dark matter \cite{BaumgartMatthew_JHEP_2009}, because the cross section in hadron collisions is very small and there is no excess of antiprotons, which has also been demonstrated by PAMELA results. 
While SUSY is not necessary to realize such a model, it is well motivated because the GeV scale is automatically generated \cite{BaumgartMatthew_JHEP_2009}.

Concerning the aforementioned ``hidden'' symmetry in the models of N.Arkani-Hamed, it is important to note several theoretical ideas which motivate a concept of a so-called ``hidden sector'' comprised of yet unobservable fields which are singlets under Standard Model $SU(3)$$\times$$SU(2)$$\times$$U(1)$ gauge symmetries 
~\cite{N.Arkani-Hamed_N.Weiner_PhysRev_2008,
N.Arkani-Hamed_N.Weiner_JHEP_2008,
BaumgartMatthew_JHEP_2009,
M.J.Strassler_2006,
M.J.Strassler_2008}. 
There is a vast amount of speculation in the literature about the nature of hidden sectors, including for example simple modifications of Standard Model ~\cite{BernabieRiccardo_2005,HoomanDavoudiasl_2005, P.Brian_Wilczek_2006}, 
Stueckelberg extensions of Standard Model~\cite{B.Kors_PhysLettB_2004, Kors:2004ri}, 
string theory variations ~\cite{D.Gross_PhysRevLett_1984}, 
models of higher dimension operators mediated by heavy states in Hidden Valleys~\cite{M.J.Strassler_2006, M.J.Strassler_K.M.Zurek_2007, HanTao_2007}, 
models with mediation via kinetic mixing \cite{Feldman:2007wj, Wells:2009kq, PhysRevD.77.087302} 
and specifically kinetic mixing in the class of dark force models discussed in \cite{N.Arkani-Hamed_N.Weiner_PhysRev_2008, BaumgartMatthew_JHEP_2009, Cheung:2009qd}, 
Higgsed dark-sector~\cite{N.Arkani-Hamed_N.Weiner_PhysRev_2008, N.Arkani-Hamed_N.Weiner_JHEP_2008, BaumgartMatthew_JHEP_2009}, 
Confined dark-sector \cite{EssigRouven_PhysRev_2009}, 
and Unparticle physics \cite{H.Georgi_PhysRevLett_2007}. 
We also discuss later in the paper generalized portals occurring due to hidden-visible sector couplings arising from both kinetic and mass mixings \cite{Feldman:2007wj, Burgess:2008ri, PhysRevD.78.075005}. Compiling a full listing all the different variants of the hidden sector (supersymmetric) models is in itself a daunting task and is beyond the scope of this work, so we recommend reader this Ref.\cite{Nath2010185, Baer:2008uu} and references therein.

\begin{figure*}
  \centering
\subfigure[]{
  \includegraphics[scale=1.05]{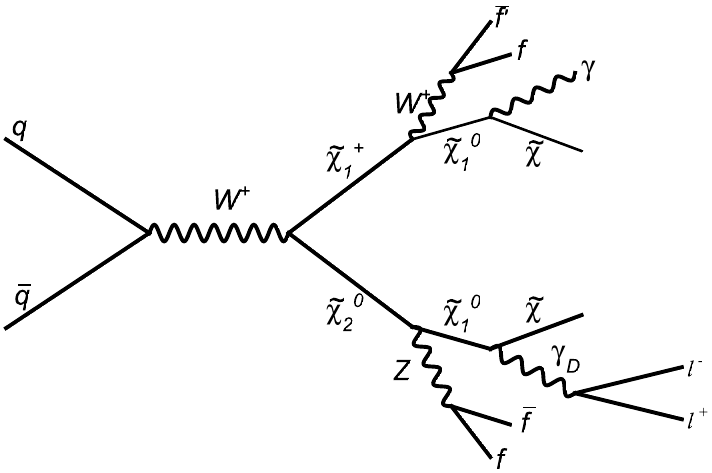}
\label{fig:DarkSector_TevatronSearch}
}
\subfigure[]{
  \includegraphics[scale=1.05]{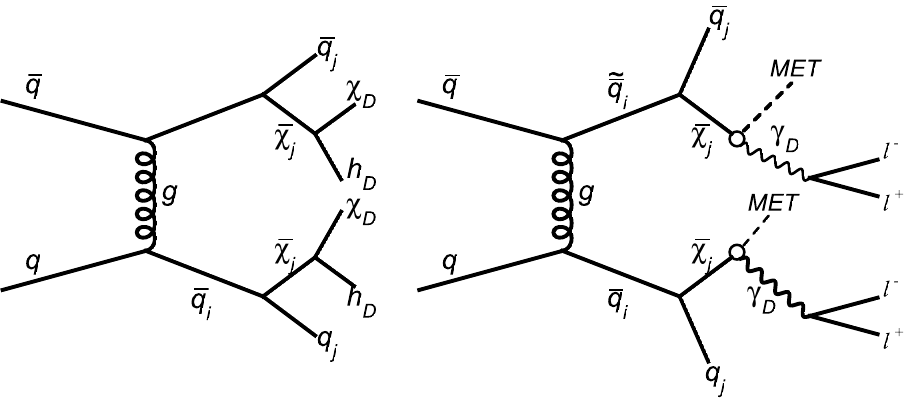}
\label{fig:DarkSector_MahsanaAhsana}
}
\subfigure[]{
  \includegraphics[scale=0.56]{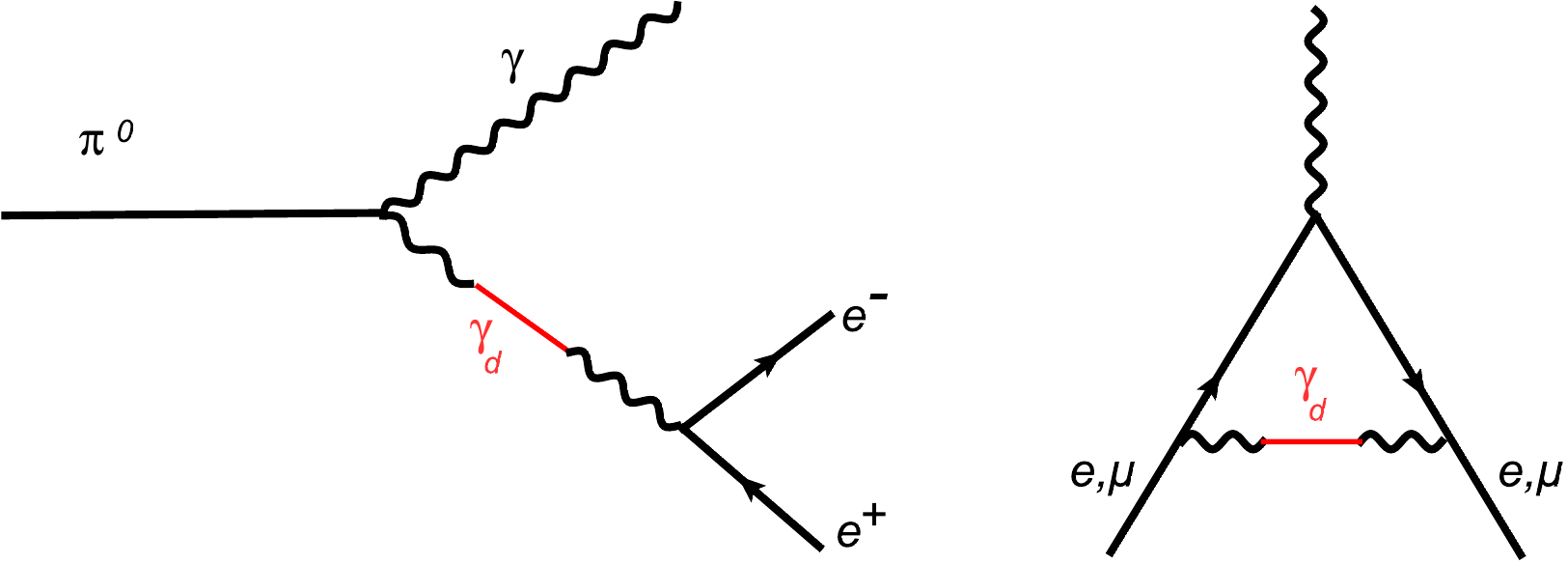}
\label{fig:DarkPhoton_WASAatCOSY_comb}
}
\subfigure[]{
\includegraphics[width=0.58\columnwidth]{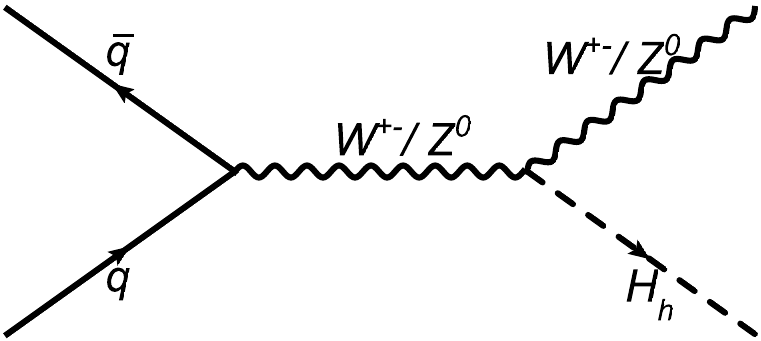}
\label{fig:DarkSector_cascade_23stepModelFullLeptonList_v2_bold}
}
\subfigure[]{
\includegraphics[width=1.93\columnwidth]{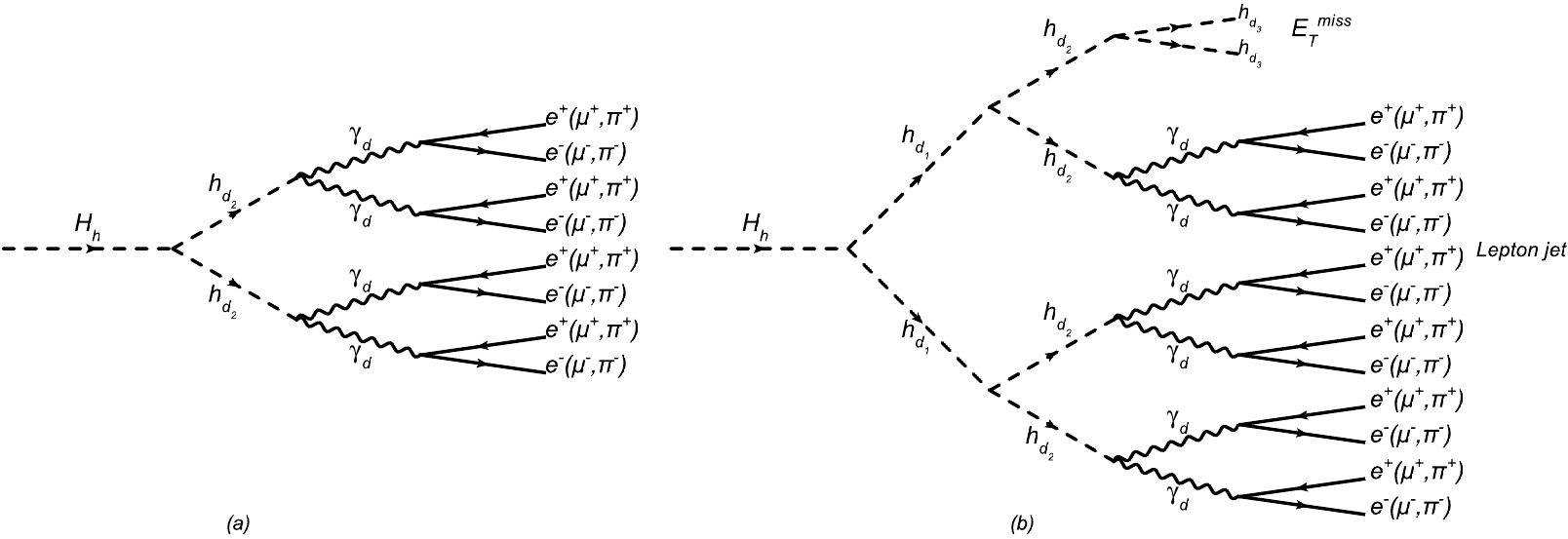}
\label{fig:DarkSector_cascade_23stepModelFullLeptonList_v2_bold}
}
\caption{
(a) One of the diagrams giving rise to the events with a photon, dark photon ($\gamma_{D}$), and large missing energy due to escaping darkinos ($\tilde{\chi}$) at the Fermilab Tevatron Collider, from \cite{Abazov:2010_ljets}. 
(b) Example of the Feynman diagrams for the production of the dark matter particles through the SUSY final state mechanism and (right) an extension of this diagram where the dark matter particles (for example dark neutralino $\chi_{D}$, and dark higgs $h_{D}$) decaying into low mass states end up with large missing energy and a boosted set of leptons via coupling to the dark photons  $\gamma_{D}$; these diagrams relate to \cite{Aad:2012qua}. 
(c) Feynman diagrams for the possible contribution of dark photons, $\gamma_{D}$, to $\pi^{0}\rightarrow e^{+}e^{-}\gamma$ and  lepton $g-2$, from \cite{Adlarson2013187}.
(d,e) Example of the Feynman diagrams for the associated Higgs boson production with $W^{\pm}$/$Z$ in $pp$ collisions, that give rise to the events with a dark photon at the LHC , from \cite{Aad:2013yqp} and \cite{Tykhonov:2013, Deliyergiyev:1995045}. In this case the Higgs subsequently decay into a light hidden/dark sector: 2-step (a) and 3-step (b), where $m_{h_{d1}}$$\textgreater$$m_{h_{d2}}$$\textgreater$$m_{\gamma_{D}}$.
}
  \label{fig:DarkSector_darkphoton}
\end{figure*}

Hidden sector models have a Higgs that decays predominantly into a light hidden sector (dark sector) either directly or through light hidden states \cite{J.T.Ruderman_T.Volansky_JHEP_02_2010, J.T.Ruderman_T.Volansky_JHEP_05_2010}. 
There are several states which are typical to this sector. Heavier dark sector states, after being produced through one of the channels (SUSY electroweak-ino production, Rare $Z$ decay, prompt dark-photon, Dark Higgs), will cascade down to lighter states; some examples are shown on Fig.\ref{fig:DarkSector_darkphoton}, and, after decaying back into SM states on measurable time scales. Similar to the QCD and QED the dark sector state can have ``dark radiations'' if the dark sector gauge coupling is not so small \cite{Buschmann:2015awa} --  or dark showering, see Fig.\ref{fig:DarkSector_darkphoton}(d,e).

Assuming that the dark photon masses are relatively small compared to the typical partonic center of mass energy at the LHC, emission of dark photons from a DM particle receives a strong enhancement in the collinear direction, so typically will be produced at the LHC with a large boost. In other words the resulting decay products are highly collimated, for example from $Z$ decay we have $\gamma$ = $m_{Z}$/$2m_{b_{\mu}}$ $\sim$ 50. This leads to a class of unique objects, so called lepton-jets \cite{N.Arkani-Hamed_N.Weiner_JHEP_2008, BaumgartMatthew_JHEP_2009, Cheung:2009su}, which are high collimated energetic leptons. 
In accordance with the branching ratios, a dark photon predominantly decays into a pair of leptons, while if it is heavy enough a dark photon also may decay into pions or kaons \cite{Buschmann:2015awa, J.T.Ruderman_T.Volansky_JHEP_05_2010}. A concrete search strategy for such ``hidden'' Higgs decays has been proposed in \cite{J.T.Ruderman_T.Volansky_PhysRevLett_2010}. 

Previous dark sector searches at the Tevatron focus on the electroweak production of charginos and neutralinos \cite{Abazov:2010_ljets}, see Fig.\ref{fig:DarkSector_darkphoton}(a), while the higher center-of-mass energy of the LHC allows to focus the search on colored SUSY production, because of higher color production cross sections, where squarks become the lightest superpartners. A search for dark photon with mass above 210 MeV, was performed by the CDF \cite{WilburScott_boost2011}; the D${\O}$ collaboration focused on dark-photon with mass 1.4 GeV \cite{Abazov:2010_ljets}, and came up with more results for $m_{\gamma_{D}}= 0.3, 0.9$, and 1.3 GeV in \cite{PhysRevLett.105.211802}. The sensitivity of that analysis to models considered in the ATLAS work is approximately two orders of magnitude lower \cite{Aad:2012qua, Aad:2013yqp, Aad:2014yea}.

There are several possible production mechanisms that involve SUSY in the dark sector, which will be discussed later in the paper. One such mechanism involves electroweak-inos \cite{Cheung:2009su} that tend to cascade down to the lightest neutralino $\chi_{0}^{1}$, producing (color) particles along the way, resulting in QCD jets in addition to lepton-jets in the model's collider signature. Another production mechanism involves squark pair production and subsequent decay \cite{BaumgartMatthew_JHEP_2009}. The ATLAS results based on this signature are shown in \cite{Aad:2012qua}. In this search, the DM particles produced through SUSY final state mechanism, where the DM particles (for example dark neutralino $\chi_{D}$, and dark higgs $h_{D}$) decaying into low mass states end up with large missing energy and a boosted set of leptons via coupling to the dark photons $\gamma_{D}$. The representation of this process is shown in Fig.\ref{fig:DarkSector_darkphoton}(b). Squarks represent the lightest supersymmetric partner (LSP) of the MSSM that decays into a quark and lighter dark sector fermion, which then cascade-decays into the lightest dark sector states, ending up with large missing energy and a boosted set of leptons or pions via coupling to the dark photons $\gamma_{D}$. The lightest dark sector fermion becomes the true LSP which escapes detection as a cold DM candidate\footnote{The simplest assumption concerning the DM is that it has no significant interactions with other matter, and that its particles have a negligible velocity as far as structure formation is concerned. Such DM is described as `cold', and candidates include the lightest supersymmetric particle, the axion, and primordial black holes.}.  

Following the hidden sector search strategy proposed in ~\cite{J.T.Ruderman_T.Volansky_PhysRevLett_2010}, the ATLAS \cite{Aad:2012qua, Aad:2013yqp} and CMS \cite{Daci:2016030, Khachatryan:2015wka} collaborations perform DM search under the assumption that DM is made up of light particles, as some of the leading theories and observational evidence suggest. They search for associated production of  ``hidden" Higgs with $W$  boson, where the domain of ``hidden/dark'' sector parameter space is considered such that Higgs decays via several steps in the dark sector (dark showering), the set of the stable scalars mimicking the lightest hidden fermions, to lepton-jets, and probe dark photon masses in the range 0.1 GeV $\textless~m_{\gamma_{D}}~\textless$ 0.2 GeV   \cite{Aad:2013yqp, Tykhonov:2013}. Additional probes of the dark photon mass have been done in the range 0.15 GeV $\textless~m_{\gamma_{D}}~\textless$ 0.5 GeV \cite{Aad:2012qua}. 

Recently ATLAS updated their limits on prompt dark photon production using 8 TeV data \cite{Aad:2015sms}. Results from an 8 TeV ATLAS search for displaced muon lepton-jets as decay products of the dark photon with mass of 0.4 GeV are shown in \cite{Aad:2014yea}. The CMS results for displaced lepton-jets are shown in \cite{CMS:2014hka, Khachatryan:2014mea}.

In addition, the DM cascade, and parton showering, in the dark sector can lead to higher multiplicities of the leptons that form these jets (possibly two, three, four, or more), however theoretical models do not give an unambiguous definition of lepton-jets. In a sense its not clear how many leptons should form a shape of such a jet. Therefore, there are too many options for model-based searches, some of which have been tested in \cite{Deliyergiyev:1995045} by analyzing the associated production of hidden Higgs with $W/Z$ boson \cite{Deliyergiyev:1995045}. This analysis  probes for dark photon masses in the range 0.1 GeV $\textless~m_{\gamma_{D}}~\textless$ 0.7 GeV  using the same model as  \cite{Aad:2013yqp}.

In comparison to hadron colliders, the low-energy $e^{+}e^{-}$ colliders offer an ideal environment to probe low-mass dark sectors \cite{V.A.Kuzmin, Nussinov198555, Chang:2009sv}. Searches for dark photons on these machines have also focused on the associated channels, where a dark photon is produced in association with a photon in $e^{+}e^{-}$ collisions and decays back to SM fermions if other dark sector states are kinematically inaccessible \cite{Aubert:2009af, PhysRevLett.113.201801}. The BaBar collaboration probes dark photon masses in the range 0.3 GeV $\textless~m_{A^{\prime}}~\textless$ 5 GeV \cite{Aubert:2009af} and later 0.02 GeV $\textless ~m_{A^{\prime}}~\textless$ 10.2 GeV \cite{PhysRevLett.113.201801}. 
Additional constraints on the dark photon parameter space from the BaBar dark Higgsstrahlung searches have been imposed in \cite{An:2015pva}.

The WASA search for short-lived dark photons in $\pi^{0}$ Dalitz decay, $\pi^{0}\rightarrow e^{+}e^{-}\gamma$, probes masses in the range 0.02 GeV $\textless~m_{U}~\textless$ 0.1 GeV \cite{Adlarson2013187}. The CERN NA48/2 collaboration also tries to search for the dark photon by focusing on the deviations in the Dalitz decays, successfully probing masses in the range 0.009 GeV $\textless~m_{A^{\prime}}~\textless$ 0.07 GeV \cite{Batley:2015lha}. The same methods have been used by the PHENIX in their search for the new boson, while in addition to $\pi^{0}$ decays that were probed by the WASA and NA48/2 they consider $\eta$ meson decays to $e^{+}e^{-}\gamma$, where an electron positron pair appears as result of the dark photon decay \cite{Adare:2014mgk}. They put upper limits  on the dark photon mixing strength for the mass range 0.03 GeV $\textless~m_{U}~\textless$ 0.09 GeV.

The KLOE did not find any evidence for dark photons in the $\phi\rightarrow\eta U$ decays, where $U$ subsequently decays to electron-positron pairs \cite{KLOE:2012}. Their resulting exclusion plot covers the mass range 0.005 GeV $\textless~m_{U}~\textless$ 0.47 GeV \cite{Archilli:2011zc}. Later they repeat the analysis on new data and put an upper limit at 90$\%$ C.L. on the ratio between dark photon coupling and the fine structure constant of $\epsilon^{2}=\alpha^{\prime}/\alpha~\textless~1.7\times10^{-5}$ for 0.03 GeV $\textless~m_{U}~\textless$ 0.4 GeV and $\epsilon^{2}=\alpha^{\prime}/\alpha~\textless~8\times10^{-6}$ for the subregion 0.05 GeV $\textless~m_{U}~\textless$ 0.21 GeV \cite{Babusci:2012cr}. The HADES result provides complementary information about the mixing $\epsilon^{2}=\alpha^{\prime}/\alpha$ of a hypothetical dark photon and set a tighter constraint than the recent WASA search \cite{Adlarson2013187} in the mass range  0.02 GeV $\textless~m_{U}~\textless$ 0.6 GeV \cite{Agakishiev2014265}.

Note, that there is no notation convention between different groups regarding this new boson, which may lead to confusion by $\gamma_{D}$,  (also $U$, $V_{\mu}$, $A^{\prime}$, $\gamma^{\prime}$ or $Z_{d}^{\prime}$), which refer to the same particle.

The results of these searches are expressed in terms of a simplified model which contains the minimal particle content and is parameterized directly in terms of the particle masses. This allows one to apply the derived methods on further searches and to set the limits to any mechanisms of communication between the hidden and visible sectors. This would help to pinpoint the mass of the DM particle in classes of hidden sector models with new confining gauge groups which are natural in a Hidden Valley, a quirk or Unparticle models and with low mass DM.

Later sections further describe DM models. The outlined classes of models lead to astrophysical predictions offering several explanations to the positron anomaly seen in the recent satellite data \cite{Adriani_PAMELA_PhysRevLett_2008, J.Chang_ATIC2005, PhysRevLett.110.141102, M.Boezi_IDM2008, Adriani_PAMELA_Nature_2008, J.Chang_Nature2008, Aharonian:2009ah, Strong_Astrophys_2005, Finkbeiner_PhysRev_2007}. Such proposals include the self-interacting DM \cite{Tulin:2012wi, Tulin:2013teo, Kaplinghat:2013xca, Kaplinghat:2015gha, Kaplinghat:2015aga}, the multicomponent DM \cite{PhysRevD.79.115002}, a boost in positrons from a Sommerfeld enhancement \cite{N.Arkani-Hamed_N.Weiner_PhysRev_2008} and a Breit-Wigner enhancement of DM annihilations \cite{PhysRevD.79.063509} (see also \cite{Feldman:2007wj}). Further, the presence of hidden sector states degenerate with the DM particle can satisfy the current relic density constraints via coannihilation effects \cite{Feldman:2006wd, Feldman:2009wv}. 

\section{Visible Signatures from Hidden Sector}
\label{VisibleSignaturesFromHiddenSectors}

This section is devoted to the review of a class of models with visible signatures due to the presence of hidden gauge symmetries. Each of these models have a hidden sector\footnote{The concept of a hidden sector is more general than supersymmetry.} consisting of particles that are completely neutral with respect to the SM gauge group, a visible sector consisting of particles of the MSSM and a communication between the hidden and the visible sectors (Fig.\ref{fig:Communication_VS_and_HS}(b)), involving the mediation by particles that comprises an additional messenger sector. There are no renormalizable tree-level interactions between particles of the visible and hidden sectors. The hidden sector models that will be discussed can be put to the test by both sets of experiments namely, direct search experiments in the underground labs and by the DM particles production via SM interaction on colliders.

The concept of the hidden sector has a long history and its modern roots lie in supersymmetry, where hidden sectors are responsible for the breaking of supersymmetry. However, the fields in the traditional SUSY hidden sectors are typically very massive. Thus while the consequences of the hidden sectors have direct bearing on the building of phenomenologically viable models, whose experimental signatures will be probed at the LHC and in dark matter experiments, the actual internal dynamics of the hidden sector are unreachable directly with colliders or cosmology, except for gravity because any DM candidate must interact gravitationally. However, more recently it has been shown that the hidden sector can give rise to unique signatures at colliders when the mass scale in the hidden sector is well below a TeV \cite{Nath2010185}, as in Hidden Valleys, Stueckelberg extensions and Unparticle models. In particular, confining dynamics in the hidden sector \cite{M.J.Strassler_K.M.Zurek_2007, H.Georgi_PhysRevLett_2007, M.J.Strassler_K.M.Zurek_2008, PhysRevLett.100.031803} give rise to exotic signatures such as high jet multiplicity events \cite{HanTao_2007} and lepton-jets \cite{N.Arkani-Hamed_N.Weiner_PhysRev_2008, J.T.Ruderman_T.Volansky_PhysRevLett_2010}, and such event multiplicities are also a feature of the models of \cite{BaumgartMatthew_JHEP_2009, Cheung:2009qd}. Thus the cascades and dynamics can become rich and complex in models with an extended hidden sector. In models of Stueckelberg mass generation and kinetic mixings, very rich event topologies arise as a consequence of gauged hidden sector vector multiplets: complex SUSY cascades and heavy flavor jet signatures from new scalars \cite{Kors:2005uz}, multi-lepton jet signals and missing energy \cite{Feldman:2007wj, Feldman:2009wv, Feldman:2006wb} as well as the possibility of mono-jet and mono-photon signatures \cite{Cheung:2007ut}; where the latter signatures also arise in the models of \cite{ Cheung:2009su, Dudas:2009uq, PhysRevD.78.095002}.

There are indeed many recent developments in hidden sector models, which include Higgs mediators, light gauged mediators and axion mediators, see e.g., \cite{ BaumgartMatthew_JHEP_2009, Wells:2009kq, Feldman:2007wj, PhysRevD.77.087302, Cheung:2007ut} and many others, as well as investigations of their phenomenological implications, see \cite{N.Arkani-Hamed_N.Weiner_JHEP_2008, Cheung:2009su, Feldman:2009wv, Cheung:2009qd, Pospelov:2008zw} and citations therein.

\subsection{Stueckelberg extensions}
\label{StueckelbergExtensions}

One of the main features of the Stueckelberg mechanism is that it allows for mass generation for a $U(1)$ vector field without the benefit of a Higgs mechanism. The $U(1)_{X}$ Stueckelberg extensions of the SM \cite{B.Kors_PhysLettB_2004}, i.e., $SU(3)_{C}$$\times$$SU(2)_{L}$ $\times$ $U(1)_{Y}$$\times$$U(1)_{X}$, involve a non-trivial mixing of the $U(1)_{Y}$ hypercharge gauge field $B^{\mu}$ and the $U(1)_{X}$ Stueckelberg field $C^{\mu}$.  There are no couplings with the visible sector fields for the Stueckelberg field $C_{\mu}$, while it may couple with a hidden sector by a gauge kinetic mixing \cite{Feldman:2007wj}. Due to LEP electroweak constraints \cite{Feldman:2006wb} these mixings, however, must be small. 

\subsection{Stueckelberg extension of MSSM}
\label{StueckelbergExtensionsMSSM}

The Stueckelberg extension of MSSM (StMSSM) is constructed from a Stueckelberg chiral multiplet mixing vector superfield multiplets for the $U(1)_{Y}$ denoted by $B$ = ($B_{\mu}$, $\lambda_{B}$,$D_{B}$) and for the $U(1)_{X}$ denoted by $C$ = ($C_{\mu}$, $\lambda_{C}$, $D_{C}$) and a chiral supermultiplet $S$ = ($\rho$+i$\sigma$, $\chi$, $F_{S}$)  \cite{Feldman:2006wb, B.Kors_JHEP_2004}.
\begin{equation} 
\mathcal{L}_{St} = \int d^{2}\theta d^{2}\bar{\theta}\left( M_{1}C+M_{2}B+S+\bar{S} \right)^{2}
\label{eq:StkExtMSSM}
\end{equation}

The Lagrangian of Eq.\ref{eq:StkExtMSSM} is invariant under the supersymmetrized gauge transformations:
$\delta_{Y}(C,B,S)$ = (0, $\Lambda_{Y}$+$\bar{\Lambda}_{Y}$, - $M_{2}\Lambda_{Y}$) 
and $\delta_{X}(C,B,S)$ = ($\Lambda_{X}$+$\bar{\Lambda}_{X}$, 0, - $M_{1}\Lambda_{X}$). 
The quantities $M_{1}$, $M_{2}$ are ``topological" input parameters of this model.  The superfield $S$ in Eq.\ref{eq:StkExtMSSM} contains a scalar $\rho$ and an axionic pseudo-scalar $\sigma$. 

A new feature of this extension is that it expands the neutralino sector of the MSSM. Also it provides an example of a model where the astrophysical implications for a wino LSP as well as a mixed Higgsino wino LSP have important effects on observables.

Dozen of testable signatures of new physics arise within the framework of the Stueckelberg extensions of the SM and of the MSSM. The minimal model produces a narrow vector resonance that is detectable in the dilepton channel and at the LHC \cite{B.Kors_PhysLettB_2004, Feldman:2007wj}. The supersymmetric extension also predicts the presence of a sharp scalar resonance in the Higgs sector (see \cite{Kors:2005uz}). 
The forward-backward asymmetry near the $Z^{\prime}$ pole can also provide a detectable signal at a linear collider \cite{Kors:2005uz}. Moreover, one should observe the mono-jet signatures if the $Z^{\prime}$ decays dominantly into the hidden sector.

The predictions in the fermionic sector are also rich with implications for DM and for the LHC. The extensions give rise to three classes of DM (a) milli-weak (b) milli-charged (c) neutralino-like with extra hidden sector degrees of freedom. Thus, the models provide a Dirac dark matter candidate \cite{Feldman:2007wj, Cheung:2007ut}, see Fig.\ref {fig:SM_to_DM_particles}(b), that can fit the WMAP data when integrating over the Breit-Wigner Poles \cite{Feldman:2007wj} and can also fit the recent astrophysical data (see Sect.\ref{INTRO_CosmicPositronPuzzle}) due the Breit-Wigner enhancement \cite{PhysRevD.79.063509} from the $Z^{\prime}$ pole. 

\section{Hidden Valleys}
\label{Hidden Valleys}

Over the past decades the weak scale became the dominant playground for developing the paradigm of DM. New symmetries giving rise to stable particles are often appearing in models that stabilize the Higgs mass at the weak scale. In many of these models the DM density, computed from the thermal relic abundance of the particles with masses at the weak scale, is consistent with the astrophysical observations. A solution to the baryon and DM coincidence puzzle arises in the subclass of the hidden sector DM models, namely Hidden Valley models. 

However, in recent years it has been realized that in these models the phenomenology can be quite distinct and difficult to find at the LHC, even if the masses of such extensions to the SM are much lighter than the weak scale. This issue was examined in the context of Hidden Valley models \cite{M.J.Strassler_K.M.Zurek_2007,  M.J.Strassler_K.M.Zurek_2008}, where a light gauged hidden sector communicates to the SM through weak scale states. One may also notice similarities and connection to ``quirk''\footnote{The $U$ and $D$ quirks are similar to quarks except they carry a new global charge that keeps one combination, $UD$, stable ($U$ and $D$ carry opposite electric charge). Therefore they charged on both hidden sector and SM gauge groups.} \cite{Kang:2008ea} and Unparticle models \cite{H.Georgi_PhysRevLett_2007}. The impact of a Hidden Valley on supersymmetric phenomenology at colliders can be significant if the LSP lies in the hidden valley sector \cite{M.J.Strassler_2006, PhysRevD.79.115002}.

In these models, states at the TeV scale are often unstable, and decay to lighter particles in the hidden sector Fig.\ref{fig:Communication_VS_and_HS}(b). This includes, for example, weak scale supersymmetric states that were previously DM candidates. Often the lightest $R$-parity odd state will reside in the hidden sector, and the MSSM dark matter candidate will decay to such a light state, modifying the DM dynamics and the freeze-out calculation \cite{ M.J.Strassler_2006}.

Might these low mass hidden sectors shatter the WIMP miracle? In most cases no. There are two reasons. First, in these hidden sectors one may naturally maintain the same annihilation rate for thermal freeze-out. The annihilation cross section needed to obtain the observed relic abundance is 
$\langle\sigma_{weak}\upsilon\rangle$ $\backsimeq$ 3 $\times$ $10^{-26}$ cm$^{3}$/s
\footnote{Thermally-averaged annihilation cross section},
logarithmically sensitive to the DM mass. 
This relation is in particular naturally obtained for weak scale DM, since $g^{4}/m_{X}^{2}$ $\backsimeq$ 3 $\times$ $10^{-26}$ cm$^{3}$/s for an $\mathcal{O}$(1) gauge coupling $g$ and weak scale DM mass $m_{X}$. 
However, if $g \ll$ 1 and $m_{X} \sim g^{2}m_{weak}$, the relation still holds for much lighter DM masses. 
This is particularly well motivated in the context of gauge mediation, where the dark hidden sector mass scale, $m_{DHS}$, is set via two loop graphs, $m_{DHS}^{2}$ $\backsimeq$ $g^{4}F^{2}/(M^{2}16\pi^{2})^{2}$$\rm{log}$$(m_{weak}/m_{DHS})$. 
Since $m_{DHS}$ scales with $g^{2}$, the WIMP miracle still holds for DM masses well below the TeV scale, if $m_{X}$ $\neq$ $m_{weak}$, see ``WIMPless miracle"\cite{PhysRevLett.101.231301}. 
In the 0.1 GeV -- 1 TeV range DM mass is naturally obtained in the case of $10^{-2}$ $\lesssim$ $g$ $\lesssim$ 0.1. 
Moreover, there are no difficulties to naturally induce even lower mass scales, such as an $\mathcal{O}$(MeV), if kinetic mixing is involved \cite{PhysRevD.77.087302}. 
However there are strong constraints obtained from electron beam-dump experiments on such MeV gauged hidden sectors \cite{Bjorken:2009mm}.

For theories with communication of the hidden sectors and the SM through kinetic mixing where supersymmetry breaking does not set the mass scale in the hidden sector, see \cite{Feldman:2007wj, Pospelov200853}. Depending on whether supersymmetry is predominantly communicated to the hidden sector through a $D$-term or gauge mediated two loop graphs, the mass scale in the hidden sector is $\sqrt{\epsilon gg_{Y}}m_{weak}$ or $\epsilon gg_{Y}m_{weak}$, where $g_{Y}$ is the hypercharge gauge coupling. For $\epsilon$ $\backsimeq$ $10^{-4}$ -- $10^{-2}$ GeV dark forces are obtained as studied in \cite{N.Arkani-Hamed_N.Weiner_PhysRev_2008, BaumgartMatthew_JHEP_2009}. For smaller $\epsilon$, lower mass dark forces may be obtained.

The second case is where solutions to the baryon-DM coincidence problem provide the observed relic abundance that is naturally obtained with DM mass well below the weak scale. In these cases a light hidden sector is, in many cases, required to reproduce the observed relic abundance. The baryon-DM coincidence is the observational similarity of the present values of the measured abundances of baryons and of DM, $\Omega_{DM}$/$\Omega_{b}$ $\backsimeq$ 5 (see \cite{0067-0049-208-2-19, Ade:2013zuv} for details), while for the standard thermal freeze-out and baryogenesis models, these two quantities are set by unrelated parameters in the model (as in the MSSM, for example, where  baryon asymmetries and the DM are set largely by CP asymmetries and DM mass, respectively). However, such coincidence could be an indication of a common origin.

Solutions to this problem often relate the asymmetric number densities of the dark matter, $n_{X}$ - $n_{\bar{X}}$ , to the baryons (or leptons), $n_{X}$ - $n_{\bar{X}}$ $\approx$ $n_{b}$ - $n_{\bar{b}}$, where the exact relations are $\mathcal{O}$(1) and depend on the particular operator transferring the asymmetries. This relation in turn implies a connection between the baryon(proton) mass and the DM mass: $m_{X}$ $\approx$ 5$m_{p}$. Here again the precise factor will depend on the particular operator transferring the asymmetry. In this case the DM is low mass and weakly coupled to the SM, residing in a Hidden Valley. 

This discussion is not meant to be in any sense a complete description of these models, but rather a broad overview of the types of hidden sectors that have been constructed. We refer the reader to the appropriate references for details on their construction. 

\subsection{Models of hidden dark matter}
\label{ModelsOfHiddenDarkMatter}

As was indicated above, low mass DM candidates may be particularly well motivated in the context of gauge mediation with kinetic mixing of a new $U(1)_{X}$ with hypercharge, as considered in \cite{N.Arkani-Hamed_N.Weiner_PhysRev_2008, BaumgartMatthew_JHEP_2009, PhysRevD.77.087302}. What happens to the dark force in the hidden sector?  As has been discussed in \cite{N.Arkani-Hamed_N.Weiner_PhysRev_2008, BaumgartMatthew_JHEP_2009, PhysRevD.77.087302}, a VEV for the dark Higgses will be induced through SUSY breaking effects, this VEV in turn will break the dark force and give it a mass set by the size of the SUSY breaking mass scale in the hidden sector, typically much lower than the TeV scale.

Hypercharge $D$-terms will induce a VEV for a dark Higgs with tan$\beta$ $\neq$ 1 induced by SUSY breaking while the supersymmetric Electroweak Symmetry Breaking necessarily has ${\rm{tan}}\beta = 1$ to ensure $D$-flatness \cite{Batra:2008rc},  through the potential
\begin{equation} 
V_{D} = \frac{g_{x}^{2}}{2} \left( \sum\limits_{i} x_{i}\vert\phi_{i}\vert^{2} - \frac{\epsilon}{g_{x}}\xi_{Y} \right)
\label{eq:darkHiggsPotential}
\end{equation}
where $\phi_{i}$ in the hidden sector scalar field with $U(1)_{X}$ charge of the Higgs $x_{i}$, $g_{x}$ the gauge coupling and $\xi_{Y}$ = -$\frac{g_{Y}}{2}c_{2\beta}\upsilon^{2}$ is the hypercharge $D$-term, with $\upsilon$ = 246 GeV and the mixing between up and down-type Higgses, $\beta$. 

The VEV for the dark Higgs induced by this potential Eq.\ref{eq:darkHiggsPotential} is
\begin{equation} 
\langle  \phi_{i} \rangle \backsimeq \left(  \frac{\epsilon\xi_{Y}}{g_{x}x_{i}}  \right)
\label{eq:darkHiggsVEV}
\end{equation}
As was outlined in the previous section, for $\epsilon$ $\backsimeq$ $10^{-4}$ -- $10^{-2}$ the dark $U(1)_{X}$ gauge boson acquires a GeV scale mass. The smaller gauge boson masses, down to MeV range, can be obtained for smaller kinetic mixings.

There is a subdominant effect relative to renormalization group effects, termed Little Gauge Mediation \cite{PhysRevD.79.115002,Morrissey:2009ur}, which communicates a soft mass to the hidden Higgs of size $m_{soft}^{hid}$ $\sim$ $\epsilon$ $m_{soft}^{vis}$ through the usual two loop gauge mediation diagrams, with messengers in the loop. The typical size of such soft terms is close to a GeV for $\epsilon$ $\sim$ $10^{-3}$ and $m_{soft}^{vis}$ $\sim$ TeV. More precisely this gives rise to the soft scalar masses, a dark Higgs mass, at the gauge messenger scale:
\begin{equation} 
m^{2}_{\phi_{i}} = \epsilon^{2}x_{i}^{2} \left( \frac{g_{x}}{g_{Y}}\right)^{2} m_{E^{c}}^{2}
\label{eq:darkHiggsMass}
\end{equation}
where $m_{E^{c}}^{2}$ is the SUSY breaking mass of the right-handed selectron \cite{PhysRevD.79.115002}. These terms are almost always important for determining the precise spectrum of the hidden sectors, particularly when the hypercharge $D$-term is zero. 

The spectrum in the hidden sector will depend on the precise matter content, however taking a simple anomaly free dark sector
\begin{equation} 
W_{d} = \lambda S \phi\bar{\phi}
\label{eq:darkMatterContent}
\end{equation}
results in one stable, $R$-odd fermion, whose mass is either $\lambda\langle\phi\rangle$ or $\sqrt{2} x_{H} g_{x}\langle\phi\rangle$.

The DM mass in these models is set by thermal freeze-out, and for some ranges of parameters and mass spectra a ``WIMPless miracle'' for dark matter in the MeV to tens of GeV mass range naturally results \cite{Morrissey:2009ur}. While in some classes of these low mass hidden sector models, thermal freeze-out naturally results in the right relic abundance.

\subsection{Low mass dark sectors as solutions to the baryon-dark matter coincidence}
\label{Low mass dark sectors as solutions to the baryon-dark matter coincidence}

Now we discuss that class of models where GeV mass states will automatically give the correct relic abundance: solutions to the baryon and DM coincidence, see Sec.\ref{Hidden Valleys} for more details. One may find a number of solutions to the baryon-DM coincidence in the literature \cite{V.A.Kuzmin}, especially in the context of technicolor \cite{Nussinov198555}. Here we will try to describe a particularly simple class of models which is termed Asymmetric Dark Matter \cite{Kaplan:2009ag}. For this class the DM candidate is not derived from models designed to stabilize the weak scale. This particular class of models fits the paradigm of the low mass hidden sector -- Hidden Valley. 

These models have been developed in order to build an effective field theory which describes the interactions between the hidden sector and the visible sector, which transfers a SM baryon or lepton asymmetry to the dark sector. There is sterileness requirement for the DM in these models, which is applied in order to reduce the number of operators which can be constructed. In particular, in the context of supersymmetry, if the sterile operators $udd$ and $LH$ -- the lowest dimension operators carrying lepton or baryon number are connected to the hidden sector containing the dark field $\bar{X}$ to transfer an asymmetry, then the superpotential takes the form
\begin{eqnarray}
&& W = \frac{\bar{X}^{2}udd}{M^{2}}\\
&& W = \frac{\bar{X}^{2}LH}{M}
\label{eq:darkOperators}
\end{eqnarray}
The second operator, for example, enforces the following relation between DM number densities 2($n_{X}$ - $n_{\bar{X}}$) = $n_{\bar{l}}$ - $n_{l}$, 
and the detailed calculation relating the lepton asymmetry to the baryon asymmetry (through sphalerons 
\footnote{The classical saddle-point solutions in a relativistic field theory, which are related to the topology of the field configuration space. Based on the classical Greek adjective $\sigma\varphi\alpha\lambda\epsilon\rho os$ ($sph\Breve{a}ler\Breve{o}s$) meaning ``ready to fall.''}
\cite{PhysRevD.30.2212_sphaleron}) consequently shows that this model predicts $m_{X}$ $\backsimeq$ 8 GeV. One may note that $\bar{X}^{2}$ which was added instead of $X$, ensures DM stability due to additional $Z_{2}$ symmetry. However, we refer to \cite{Chang:2009sv} in cases where $R$-parity is utilized instead to stabilize the DM.

After transferring the SM baryon or lepton asymmetry to the dark sector, the symmetric part of the DM (which is much larger than the asymmetric part, $n_{X}$ + $n_{\bar{X}}$ $\gg$ $n_{X}$ - $n_{\bar{X}}$) must annihilate, leaving only the asymmetric part. There are a variety of mechanisms to do this, but the difficulty here is having a mechanism which is efficient enough to annihilate away the whole of the symmetric part through $X\bar{X}$ $\rightarrow$ $SM$. The cross-section for such a process can be defined through a dimension six operator as
\begin{equation} 
\sigma \upsilon = \frac{1}{16\pi}\frac{m_{X}^{2}}{M^{\prime 4}}
\label{eq:DimensionXsection}
\end{equation}
In order to reduce the DM density to its asymmetric component this cross-section must be bigger than approximately 1 pb. This requirement implys a rather severe constraint for any new electroweak state coupling to the SM states, $M^{\prime}$ $\lesssim$ 100 GeV.

Here a confining gauge interaction in the hidden sector, which introduces a mass gap into the theory, can be a useful tool. If the symmetric and asymmetric bound states of elementary dark sector fermions form the DM, then the symmetric states may decay through the same dimension six operators, while the asymmetric states would remain stable \cite{Zurek:2010xf}. 

\subsection{Dark sectors with confinement}
\label{Dark sectors with confinement}

The constituents of the DM bound states may carry electroweak charges. This possibility was recently considered in a framework of confinement models \cite{Kribs:2009fy}.
One may note that these models bear some similarity to the models constructed in the context of technicolor \cite{Nussinov198555}. 
The new defining characteristic of this hidden sector model is the presence of a new non-Abelian gauge group which confines at a low scale. The DM candidate is a charge-neutral composite of electroweak charged, weak scale mass, ``quirks''. That is to say, analogous to the proton, the DM is a composite dark baryon. 
The low mass dark glueballs reside in the hidden sector, while the DM constituents are themselves heavy weak scale fields and act as the connectors between the SM and dark glue sector. 

These constituents can be processed by sphalerons, because  they are electroweak-charged. In particular, the sphalerons will violate some linear combination of baryon, lepton and dark baryon number, $DB$. Thus an asymmetry in baryon and lepton numbers (produced from some leptogenesis or baryogenesis mechanism) will be converted to an asymmetry in $DB$. The $DB$ asymmetry then in turn sets the DM relic density. Since the DM mass is around the mass of the weak scale quirk constituents, there must be a Boltzmann suppression in $DB$ to achieve the observed baryon-DM coincidence \cite{0067-0049-208-2-19, Ade:2013zuv}. This can be naturally achieved when the sphalerons  decouple just below the DM mass \cite{PhysRevD.30.2212_sphaleron, PhysRevD.36.581_sphaleron}:
\begin{equation} 
\Omega_{DM} \sim \frac{m_{DM}}{m_{p}}e^{-m_{DM}/T_{sph}}\Omega_{b}
\label{eq:darkMatterMass}
\end{equation}
where $T_{sph}$ is the sphaleron decoupling temperature \cite{PhysRevD.36.581_sphaleron}, and the exact proportions are worked out in \cite{Kribs:2009fy}. 

These models have also effectively been used to achieve the mass splittings necessary to realize the inelastic \cite{PhysRevD.79.043513, PhysRevD.79.115011} and exciting DM scenarios \cite{Finkbeiner_PhysRev_2007, Alves_2009nf}. 
The DM in the dark sectors with confinement is again at the weak scale composite with the confinement scale of the gauge group binding the constituents in the 100 keV - MeV range. The result is mass splittings between the DM ground state and excited states set by the confinement scale, and these mass splittings are phenomenologically of the size to fit DAMA \cite{BernabieRiccardo_2008} and INTEGRAL \cite{Strong_Astrophys_2005} observations through the excitation of the DM ground state to one of the higher states, which then decays back to the ground state, producing $e^{+}e^{-}$ or resulting in an inelastic scattering of DM off nuclei.

\section{GeV dark sector at the LHC}
\label{Probing the GeV dark sector at the LHC}

Dark matter can carry GeV$^{-1}$ scale self-interactions \cite{Spergel:2000}, but considering the thermal evolution of the dark and visible sectors in the early Universe  may shift the scale to TeV$^{-1}$ \cite{Foot:2014osa}. The self-interactions of DM have important implications for the formation and evolution of structure, from dwarf galaxies to clusters of galaxies \cite{Tulin:2012wi, Tulin:2013teo, Kaplinghat:2015aga}. The equilibrium solutions in these self-interacting DM models for the dark matter halo density profile, including the gravitational potential of both baryons and DM, has been obtained in \cite{Kaplinghat:2013xca, Kaplinghat:2015gha}.

Motivated by astrophysical observations (see Sect.\ref{INTRO_CosmicPositronPuzzle}), it has been proposed in Ref.\cite{N.Arkani-Hamed_N.Weiner_PhysRev_2008, Pospelov200853}) that electroweak scale DM ($m_{DM}$ $\sim$ TeV) has GeV$^{-1}$ range self-interactions. This sector generically also couples to the SM states in order to account for the excesses in the cosmic ray observations. In the same time, in order to satisfy the experimental constraints, such couplings (the ``portal'') are expected to be tiny. A review of more specific model buildings for the dark sector can been found in \cite{Nath2010185}. One may also note that this class of models can be regarded as a distinct possibility of the Hidden Valley scenario 
\cite{ M.J.Strassler_K.M.Zurek_2007, M.J.Strassler_K.M.Zurek_2008}.

Different choices of the dark sector gauge interactions, $G_{D}$, and the portal to the SM have been considered in \cite{PhysRevD.79.075008}. In the following, we will focus on the portal which is generated by kinetic mixing between an $U(1)_{Y}$ factor of $G_{D}$ and the hypercharge $U(1)_{Y}$. Let's discuss the most relevant part of the Lagrangian from which the most generic signals can be derived. The kinetic mixing can be parameterized as \cite{Cheung:2009su}
\begin{equation} 
\begin{split} 
\mathcal{L}_{gauge ~\rm{mix}} &= -\frac{1}{2}\epsilon_{1} b_{\mu\nu}A^{\mu\nu} -\frac{1}{2}\epsilon_{2} b_{\mu\nu}Z^{\mu\nu} \\
&= -\frac{1}{2}\epsilon_{1}^{\prime} b_{\mu\nu}B^{\mu\nu} -\frac{1}{2}\epsilon_{2}^{\prime} b_{\mu\nu}W^{\mu\nu}_{3},
\label{eq:DS_KineticMixing_parameterization}
\end{split} 
\end{equation}
where $b_{\mu\nu}$ denotes the field strength for the dark gauge boson and $\epsilon_{1,2}$ and $\epsilon_{1,2}^{\prime}$ are related by the Weinberg angle. In particular, when only $\epsilon_{1}^{\prime}$ is present, we have 
$\epsilon_{1}$ = $\epsilon_{1}^{\prime}$ cos$\theta_{W}$ and 
$\epsilon_{2}$ = $\epsilon_{1}^{\prime}$ sin$\theta_{W}$. 
The fields $B^{\mu\nu}$ and $W^{\mu\nu}_{3}$ are replaced by the physical particles $Z^{\mu\nu}$ and $A^{\mu\nu}$ through the rotation
\begin{equation}
\begin{split}
Z^{\mu\nu}= {\rm{cos}}\theta_{W}W^{\mu\nu}_{3}-{\rm{sin}}\theta_{W}B^{\mu\nu},\\
A^{\mu\nu}= {\rm{cos}}\theta_{W}B^{\mu\nu}_{3}+{\rm{sin}}\theta_{W}W^{\mu\nu}_{3}.
\end{split}
\label{eq:DS_hypercharge_replacement}
\end{equation}
There is also an identical mixing between the gauginos in the supersymmetric scenarios \cite{Cheung:2009su}
\begin{equation} 
\mathcal{L}_{gaugino ~\rm{mix}} = -2i\epsilon_{1}^{\prime}\tilde{b}^{\dag}\bar{\sigma}^{\mu}\partial_{\mu}\tilde{B}-2i\epsilon_{2}^{\prime}\tilde{b}^{\dag}\bar{\sigma}^{\mu}\partial_{\mu}\tilde{W}_{3}+\rm{h.c.}
\label{eq:DS_gauginos_Mixing}
\end{equation}
The kinetic mixings can be removed from by appropriate field redefinitions, which lead to the portal couplings \cite{Cheung:2009su}
\begin{equation} 
\mathcal{L}_{\rm{portal}} = \epsilon_{1} b_{\mu}J_{EM}^{\mu} +\epsilon_{2} Z_{\mu}J_{b}^{\mu} +\epsilon_{1}^{\prime} \tilde{B}\tilde{J}_{\tilde{b}} +\epsilon_{2}^{\prime} \tilde{W}_{3}\tilde{J}_{\tilde{b}},
\label{eq:DS_portal}
\end{equation}
\begin{eqnarray}
&&J_{b}^{\mu} = g_{d}\sum\limits_{i}q_{i} \left(  i\left( h_{i}^{\dag}\partial^{\mu}h_{i} - h_{i}\partial^{\mu}h_{i}^{\dag}\right)   + {\tilde{h}}_{i}^{\dag}\bar{\sigma^{\mu}}\tilde{h}_{i} \right),\\
&&\tilde{J}_{\tilde{b}} = -i \sqrt{2}g_{d} \sum\limits_{i}q_{i}\tilde{h}_{i}^{\dag}h_{i},
\label{eq:DS_currents}
\end{eqnarray}
where $J_{EM}$ is the SM electromagnetic current. $J_{b}$ contains dark scalar and dark fermion bilinears, and $\tilde{J}_{\tilde{b}}$ contains mixed dark scalar-fermion bilinears. $\tilde{B}$, $\tilde{W}_{3}$ stays for MSSM gauginos, a $\tilde{b}$ -- dark bino, $b_{\mu}$ denote the dark photon.

The range $\epsilon_{i}$ $\backsimeq$ $10^{-4}$ -- $10^{-3}$, which satisfies all the constraints (for recent studies, see \cite{Pospelov:2008zw, Reece:2009un} and references therein) implies dark photons with very short lifetimes \cite{J.T.Ruderman_T.Volansky_PhysRevLett_2010}, which leads to the prompt decays of the dark photon \cite{Aad:2012qua, Aad:2013yqp, Khachatryan:2015wka} (see also for more details on the prompt lepton-jet  search \cite{Deliyergiyev:1995045, Tykhonov:2013, Namasivayam:2053200}). For sufficiently small values of $\epsilon_{i}$ (less than $10^{-4}$), the dark photon lifetime can be sizable, leading to so-called displaced decay vertices observable in the laboratory frame \cite{Aad:2014yea, CMS:2014hka, Khachatryan:2014mea}. Therefore, most of the recent LHC searches focus on the simplest case $G_{D}$ = $U(1)_{Y}$ and try to investigate both options, which encapsulates the main features of dark sector phenomenology \cite{BaumgartMatthew_JHEP_2009, Cheung:2009qd, Cheung:2009su}. The new features can be highlighted from a more complicated dark sector.

\section{Summary}
\label{Dark Sectors Conclusions}

We have conducted a review of progress on the search for the hidden/dark photon, and on the main achievements from recent experimental investigation in this field. We are beginning to learn that the dark sector could be complex -- it may not simply be a single, stable, weakly interacting particle. There may be multiple resonances in the hidden sector with an array of new forces that govern their interactions, from confining gauge groups to a dark $U(1)$. And this new dynamic need not reside at the weak scale, which opens new avenues for exploration.  

The models mentioned in this paper address the recent astrophysical observations, showing an anomalous excess of cosmic-ray leptons, by proposing a variety of settings that impact the hidden sector and which give rise to a plethora of new physics signatures both in direct DM experiments and at the LHC, which we search for, see Fig.\ref{fig:SM_to_DM_particles}. 
Many recent theories suggest that DM is made up of previously unknown particle(s) on the scale of weak interactions \cite{N.Arkani-Hamed_N.Weiner_PhysRev_2008}. 
Specifically, the classes of hidden sector models with low mass DM, which can arise via kinetic mixings, as well as via asymmetric DM models, and dark sectors with new confining gauge groups. A couple of mechanisms for communication between the hidden and visible sectors, aside from by gravity, have been outlined. This communication could be realized via $U(1)$ gauge fields in the hidden sector which mix with the gauge fields in the visible sector via kinetic mixings or via mass mixing by the Stueckelberg mixing mechanism, or via higher dimensional operators.

It has been also shown that the recently discovered Higgs boson \cite{ATLAS:2012ae, ATLAS_HiggsObservation_2012,  Chatrchyan:2012xdj} may act as a link between particles we are familiar with and other particles that have so far avoided detection, such as DM \cite{Aad:2013yqp, Deliyergiyev:1995045, Tykhonov:2013}. Searches for new physics that are not tuned on a specific model can remain sensitive to new physics that may not be well described by known theories, and may continue to probe a wide variety of New Physics processes even if some of them are later excluded. Furthermore, the results of such model-independent searches can later be re-interpreted in the context of new models as they are proposed.

The searches reviewed in this paper seek to further test the SM by looking for deviations in the associated neutral mesons decays \cite{Adlarson2013187, Batley:2015lha, Adare:2014mgk, KLOE:2012,Archilli:2011zc,Babusci:2012cr,Agakishiev2014265}, for anomalous production of the tightly collimated prompt \cite{Aad:2012qua, Khachatryan:2015wka} or displaced lepton pairs \cite{Aad:2014yea}, and also by looking for the anomalous production of events with gauge bosons and at least two prompt leptons in the final state, $e$ and $\mu$ lepton flavors \cite{Aad:2013yqp, Deliyergiyev:1995045, Tykhonov:2013}. As a signature-based search, specific models of new physics are only considered to benchmark the performance of the selection criteria. If the dark photon decays into quarks \cite{J.T.Ruderman_T.Volansky_JHEP_05_2010, Buschmann:2015awa} it is for all intents and purposes invisible to ATLAS and CMS, if the QCD is not suppressed by some mechanism. For the heavy dark photons the aim would be, instead of a possibly hopeless search for pion pairs, to find jets which have peculiar electromagnetic vs hadron fractions \cite{J.T.Ruderman_T.Volansky_PhysRevLett_2010} -- because there would be a mixture of dark photons decaying into $e/\mu$ and $\pi$'s, which is  probably quite different to regular QCD. Hence, it is still an open issue and needs further investigation in the lepton-jet analysis approach. Efforts with smaller-scale lab experiments to search for much lighter hidden photons, where the decay channel of dark photon to lepton pair is forbidden, should also receive some attention in this scope, such as the laser polarization \cite{Redondo:2008aa, Zavattini:2005tm,Chen:2006cd}, LSW experiments \cite{Ahlers:2007qf, Ruoso:1992} and the conversion experiments, ``helioscopes'' \cite{PhysRevD.47.3707, An:2013yua, An:2014twa}.

What can we anticipate for the future? Current data from the direct experiments \cite{Akerib:2005zy, Ahmed:2008eu, PhysRevLett.100.021303, PhysRevLett.93.211301, PhysRevLett.101.091301,BernabieRiccardo_2005, BernabieRiccardo_2008, BernabieRiccardo_2010, Hooper:2011hd, Sinervo:2002sa, Aalseth:2014eft, CoGeNTColl_PhysRevLett.106.131301, Angloher:2011uu, Akerib:2013tjd} and as astrophysical experiments \cite{Adriani_PAMELA_PhysRevLett_2008, J.Chang_ATIC2005, Fermi_GLAST, Aguilar2007145, PhysRevLett.110.141102} can potentially be explained within several DM frameworks \cite{Adriani_PAMELA_PhysRevLett_2008, J.Chang_ATIC2005, PhysRevLett.110.141102, M.Boezi_IDM2008, Adriani_PAMELA_Nature_2008, J.Chang_Nature2008, Aharonian:2009ah, Strong_Astrophys_2005, Finkbeiner_PhysRev_2007}, see Fig.\ref{fig:SM_to_DM_particles}. 

However,  the smaller-scale lab experiments \cite{Ahlers:2007qf, Zavattini:2005tm,Chen:2006cd, Ruoso:1992, Redondo:2008aa, An:2013yua, An:2014twa}, at the high energy colliders \cite{Aad:2013yqp, Aad:2012qua, Daci:2016030,  PhysRevLett.113.201801, Aubert:2009af, Adare:2014mgk, Sezen:2010} and low energy colliders \cite{Adlarson2013187, Batley:2015lha, Babusci:2012cr, Agakishiev2014265, PhysRevLett.106.251802} have not yet observed any hint of a dark/hidden photon.  Without a clear positive signal, the dark photon dark matter hypothesis stands out as increasingly interesting and deserves serious attention. 

Along the way are many opportunities to detect other DM candidates, which are also strongly motivated by embedding of the Standard Model into more fundamental theories. In fact, a number of high-scale and small-scale experiments at the high and low-energies are actively searching for these elusive particles, complementing searches for physics beyond the Standard Model at the high-energy frontier. The next Run of the LHC experiments, which include scaled-up versions of the existing techniques as well as innovative concepts, together covering a huge unexplored parameter space, will continue to place tighter constraints on the exotic signatures. Discovery of DM candidate(s) would have a tremendous impact on our understanding of fundamental physics, and astrophysics, and may shed light on the dark universe.

\section{ACKNOWLEDGMENTS}
\label{ACKNOWLEDGMENTS}

The author wish to thank the unknown referees for their feedback and discussion, for bringing relevant references to my attention and to help me make this paper more clear and consistent. Moreover, the author also highly appreciates support of the Experimental Physics Department at the Josef Stefan Institute in Slovenia and Department of High Energy Nuclear Physics of the Institute of Modern Physics in China, where he completed this paper

\end{article}








\end{document}